\documentclass[lettersize, journal]{IEEEtran}
\usepackage{float}
\usepackage{caption}
\usepackage{color}

\usepackage{cite}
\usepackage{amsmath,amssymb,amsfonts}
\usepackage{amsthm}
\usepackage{graphicx}
\usepackage{textcomp}
\usepackage{xcolor}
\usepackage{svg}
\usepackage{subcaption}

\usepackage{mathtools}

\usepackage[ruled,vlined]{algorithm2e}
\usepackage{algpseudocode}

\usepackage{stfloats}

\newtheorem{theorem}{Theorem}

\newtheorem{lemma}{Lemma}

\newtheorem{corollary}{Corollary}

\allowdisplaybreaks
\makeatletter
\def\ScaleIfNeeded{%
\ifdim\Gin@nat@width>\linewidth \linewidth \else \Gin@nat@width
\fi } \makeatother

\begin{document}

\title{Exploiting Movable-Element STARS for Wireless Communications}
\author{Jingjing Zhao, 
Quan Zhou, 
Xidong Mu,
Kaiquan Cai,
Yanbo Zhu,
and Yuanwei Liu,~\IEEEmembership{Fellow,~IEEE}
\thanks{J. Zhao, Q. Zhou, K. Cai, and Y. Zhu are with the School of Electronics and Information Engineering, Beihang University, Beijing, China. (e-mail:\{jingjingzhao, quanzhou, caikq, zhuyanbo\}@buaa.edu.cn). }
\thanks{X. Mu is with the Centre for Wireless Innovation (CWI), Queen's University Belfast, Belfast, BT3 9DT, U.K. (e-mail: x.mu@qub.ac.uk).}
\thanks{Y. Liu is with the Department of Electrical and Electronic Engineering, the University of Hong Kong, Hong Kong, China (e-mail: yuanwei@hku.hk).}
}

\maketitle
\begin{abstract}
A novel movable-element enabled simultaneously transmitting and reflecting surface (ME-STARS) communication system is proposed, where ME-STARS elements positions can be adjusted to enhance the degress-of-freedom for transmission and reflection. For each ME-STARS operating protocols, namely energy-splitting (ES), mode switching (MS), and time switching (TS), a weighted sum rate (WSR) maximization problem is formulated to jointly optimize the active beamforming at the base station (BS) as well as the elements positions and passive beamforming at the ME-STARS. An alternative optimization (AO)-based iterative algorithm is developed to decompose the original non-convex problem into three subproblems. Specifically, the gradient descent algorithm is employed for solving the ME-STARS element position optimization subproblem, and the weighted minimum mean square error and the successive convex approximation methods are invoked for solving the active and passive beamforming subproblems, respectively. It is further demonstrated that the proposed AO algorithm for ES can be extended to solve the problems for MS and TS. Numerical results unveil that: 1) the ME-STARS can significantly improve the WSR compared to the STARS with fixed position elements and the conventional reconfigurable intelligent surface with movable elements, thanks to the extra spatial-domain diversity and the higher flexibility in beamforming; and 2) the performance gain of ME-STARS is significant in the scenarios with larger number of users or more scatterers. 
\end{abstract}
\begin{IEEEkeywords}
Simultaneously transmitting and reflecting surface, movable element, position optimization, beamforming.
\end{IEEEkeywords}
\section{Introduction}
With the proliferation of high-volume data transmission in the wireless communication networks, extensive research efforts have been devoted to more spectrum-efficient technologies, such as the multiple-input multiple-output (MIMO)~\cite{6736761,6375940}. Although the MIMO technique can provide substantial beam gain by exploiting the degrees of freedom (DoFs) in the spatial domain, the growing number of radio frequency (RF) chains lead to considerable hardware costs and energy consumption, especially in high frequency bands~\cite{8241348}. As a remedy, reconfigurable intelligent surfaces (RISs) have emerged as a promising solution~\cite{zeng2020reconfigurable,8910627}, which effectively enhance the signal strength via the phase response of a set of low-cost passive elements. Compared to the conventional antenna system with active hardware components, RISs that just passively adjust the propagation of incident signals are more economical and environmentally friendly, and thus can be densely deployed in wireless networks. However, the reflecting-only RIS requires the transmitter and receiver to be located on the same side of the RIS, which results in ``half-space" coverage and therefore restricts the deployment flexibility. To overcome this limitation, the simultaneously transmitting and reflecting surface (STARS)~\cite{mu2021simultaneously,xu2021star,10718344} is a promising technology to support the transmission and reflection simultaneously and facilitate the ``full-space" coverage. As such, the users on both sides of the STARS can benefit from the beam gain brought by the reconfigured propagation environment~\cite{9849460,10550177}. 

Note that conventional MIMO/RIS/STARS technologies generally employ fixed position antennas (FPAs), where the spacing among antennas is commonly set to half of the wavelength. Such discrete antenna deployment can not fully exploit the spatial DoFs and thus results in array gain loss within the antenna region. Although the antenna selection (AS) technique can further improve the utilization of the spatial diversity~\cite{1341263}, where antennas with preferable channel conditions have to be selected from a dense array for enhancing the communication performance, it results in high hardware cost due to the deployment of a massive number of antennas. To tackle this issue, the position-adjustable antenna (PAA) technologies, such as fluid antenna (FA)~\cite{9264694,9650760} and movable antenna (MA)~\cite{10318061,10354003}, have been recently studied for further enhancing the system performance. Specifically, the antennas positions can be flexibly changed within a confined region in the order of several to tens of wavelengths, in the purpose of constructing favorable channel conditions. The PAA technologies can efficiently exploit the channel variation in the continuous spatial domain, thus bringing new opportunities to the MIMO and RIS/STARS communication systems for enhancing the channel power gains or mitigating the interference. 

% Given the aforementioned benefits, the studies on MA-enabled MIMO communications have attracted some research contributions recently, while the MA-enabled RIS studies are still in its fancy. In the following, we provide the review of literatures regarding the MA-enabled communications. 
\subsection{Related Works}
Compared to the conventional FPA system, the PAA system possesses the additional spatial-domain variables, i.e., the antenna positions, to be optimized for performance improvement. The authors of~\cite{9264694} proposed the concept of FA for freely switching the position of an antenna to a set of candidate ports over a fixed-length line space, with the aim of picking up the strongest channel gain. Moreover, the authors of~\cite{9650760} studied the FA-assisted multiuser communications, where the antenna at each user was moved to the position with deep fade of the interference and favorable power gain of the desired signals.  
The authors of~\cite{10318061} proposed a novel field-response channel model for the MA system and analyzed the maximum channel gain in both deterministic and stochastic channels. Moreover, considering the multiuser communications scenario where each user is deployed with a single MA, the multiple access gain was evaluated in~\cite{10354003} by jointly optimizing the antenna positions, the user transmit powers and the BS combining matrix. Considering that the beam pattern of a MIMO system is affected not only by the beamforming weights but also by the antenna positions, the antenna position optimization was exploited in~\cite{10243545} for enhancing the channel capacity. With the objective of the minimum achievable rate maximization in the multiuser communications where the BS is employed with a MA array, the particle swarm optimization method was leveraged in~\cite{xiangyu} for solving the joint antenna position and beamforming optimization problem. Incorporating the non-orthogonal multiple access technique, the authors in~\cite{MA-NOMA} focused on the joint power allocation and antenna positions optimization via an alternating optimization (AO) scheme underpinned by the successive convex approximation (SCA) method. Thanks to the more flexible beam pattern design by exploiting the spatial variation, the MIMO system enhanced by the PAA technologies is expected to be more robust against interference. Therefore, the enhanced multi-beam forming with a linear MA array was investigated in~\cite{10382559} for striking the trade-off between the beam gain maximization over desired directions and the interference minimization over undesired directions. 

Some recent studies have started to explore the integration of RIS with PAA technologies. 
Specifically, the authors of~\cite{10794591} proposed a FA enabled joint transmit and receive index modulation transmission scheme for the RIS-assisted millimeter-wave communications. In~\cite{10716282}, the system outage probability was derived for the RIS-aided communication system involving the receiver with a single FA. 
Considering the downlink transmission from a BS deployed with a MA array to a single-antenna user with the aid of a RIS, the authors of~\cite{Xinwei} aimed to maximize the signal-to-noise ratio (SNR) by jointly optimizing the BS/RIS
active/passive beamforming and the antenna position. In a multiuser communication system with a MA enabled BS and a RIS with FPAs, a fractional programming-based iterative algorithm was proposed in~\cite{FP} for the sum-rate maximization. Furthermore, a MA-aided integrated sensing and communication system was investigated in~\cite{MA-ISAC}, where an RIS was employed to enhance both the communication and sensing performance. 
However, the works \cite{FP,MA-ISAC,Xinwei} all assumed that the RISs were equipped with FPAs and the spatial DoFs was not been fully exploited at the RISs side. Therefore, some recent works\cite{10430366,MA-RIS-YAN,MA-RIS-Wu} have started to explore the benefits of deploying MA enabled RISs, where the passive elements can be moved to desired positions for superior incoming and outgoing channels. In~\cite{10430366}, the MAs were employed at the RISs to eliminate the phase distribution offset across different cascaded source-element-destination channels when the discrete phase shifts were considered. The authors of~\cite{MA-RIS-YAN} evaluated the impact of transmit power and number of elements on the outage probability performance of a MA enabled RIS structure. However, the movement of the elements were restricted to the one-dimension (1D) region in both~\cite{10430366} and~\cite{MA-RIS-YAN}. Considering the RIS is equipped MAs that can move on the two-dimensional (2D) surface, the authors of~\cite{MA-RIS-Wu} proposed a product Riemannian manifold optimization method for the joint design of the beamforming and the elements positions, and demonstrated that enabling the movement of elements in RIS leaded to higher performance gains compared to integrating MAs with BS.
\subsection{Motivations and Contributions}
Although it has been demonstrated that PAA technologies can bring in significant performance gain to the MIMO and RIS communication system, the integration of PAA with STARS has not been well studied. In this work, we propose a new communication paradigm where the multiuser downlink transmission is assisted by a STARS deployed with movable elements (MEs), namely the ME-STARS. Note that, compared to the PAA enabled MIMO/RIS communications, exploring the full potential of the ME-STARS introduces new challenges. Firstly, different from PAAs that are deployed at the transmitter/receiver sides, the STARS MEs positions make an impact on the BS-STARS-user cascaded channel, which makes the position optimization more complicated. 
Secondly, compared to the RIS with PAAs, the STARS MEs positions can affect both the transmission and reflection channels. Thirdly, since the STARS transmission and reflection beamforming are coupled together, the joint MEs positions and beamforming optimization becomes intractable to handle. These challenging issues motivate us to contribute this work to fully reap the benefits of the ME-STARS.

Against the above background, we explore the ME-STARS-aided wireless communications and investigate the joint MEs positions and active/passive beamforming design for three STARS operating protocols, i.e., energy splitting (ES), mode switching (MS), and time switching (TS). The main contributions of this work are summarized as follows:
\begin{itemize}
    \item We propose a ME-STARS-aided downlink multiuser communication system, where the spatial-domain DoFs is further enhanced via the flexible configuration of the STARS MEs positions. By characterizing the cascaded BS-STARS-user channel power gain as a function of the MEs positions, we formulate the joint MEs positions and beamforming optimization problems for ES, MS, and TS protocols, with the aim of maximizing the weighted sum rate (WSR). 
    \item We propose an AO based algorithm to solve the highly-coupled non-convex optimization problem for ES, where the original problem is decomposed into three subproblems. For the MEs positions optimization, the penalty method is first adopted to transform the constrained problem to an unconstrained one. Afterwards, the gradient descent algorithm (GDA) is developed to find the local-optimal solution. For the active and passive beamforming, the weighted minimum mean square error (WMMSE) and SCA algorithms are invoked to obtain the stationary solutions. We further demonstrate that the proposed algorithm can be extended to solve the problems formulated for MS and TS protocols.
    \item Numerical results unveil that 1) the ME-STARS can facilitate considerable performance gain compared to the conventional STARS with fixed-position elements (FPE-STARS) for all operating protocols; 2) the ME-STARS shows superior performance than the RIS with MEs (ME-RIS) for both ES and MS, while the ME-STARS outperforms the ME-RIS for TS only when the number of users is limited; and 3) the superiority of the ME-STARS is significant in the scenario with larger number of users or more scatterers.

\end{itemize}
\subsection{Organization and Notations}
The remaining structure of this paper is arranged as follows. In Section~II, the system model is first introduced, which is followed by the ME-STARS-aided communications channel model characterization and the WSR maximization problem formulation. In Section III, the AO-based iterative algorithms are developed to address the resulting non-convex problems for three STARS operating protocols. Section IV presents the simulation results, and Section V concludes the paper. 

$\textit {Notations}$: Scalars, vectors, and matrices are denoted by italic letters, bold-face lower-case, and bold-face upper-case, respectively. $\mathbb{C}^{N\times M}$ denotes the set of $N\times M$ complex-valued matrices. Superscripts $(\cdot)^*, (\cdot)^T, (\cdot)^H$, and $(\cdot)^{-1}$ denote the conjugate, transpose, conjugate transpose, and inversion operators, respectively. $|\cdot|$ and  $\left\|\cdot\right\|$ denote the  determinant and Euclidean norm of a matrix, respectively. $\text{Tr}\left(\cdot\right)$, $\left\|\cdot\right\|_F$, and $\text{vec}\left(\cdot\right)$ denote the trace, Frobenius norm, and vectorization of a matrix,  respectively. $[\cdot]_{m,n}$ denotes the $(m,n)$-th element of a matrix. $\mathbf{1}_{{N}}$ denotes the all-one row vector with length $N$. 
$\mathbb{E}$ denotes the expectation operator. $\circ$ denotes the Hadamard multiplication. All random variables are assumed to be \textit{zero} mean. 

\section{System Model and Problem Formulation}
In this section, we present the system model of a ME-STARS-aided downlink multiuser communication system and formulate the joint MEs positions and beamforming optimization problem for the ES, MS, and TS protocols.
\subsection{System Description}
\begin{figure}
        \centering
        \includegraphics[scale=0.16]{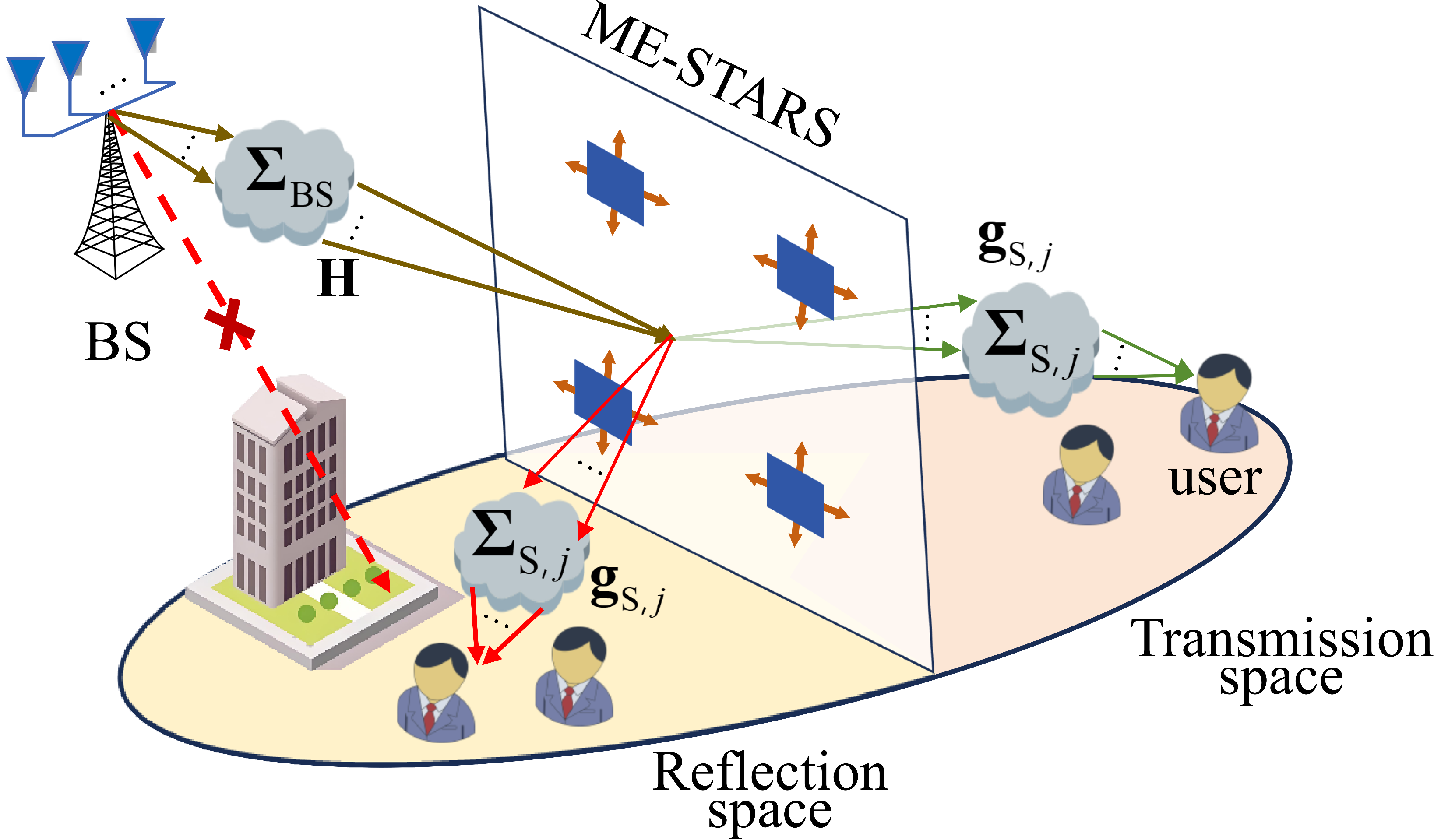}
        \caption{The ME-STARS-aided downlink multiuser communication system, where STARS MEs can move flexibly within a confined region.}
        \label{fig:system_model}
\end{figure}

As shown in Fig.~\ref{fig:system_model}, we consider a narrowband ME-STARS aided downlink multiuser communication system, which consists of a base station (BS) equipped with $M$ FPAs located in the $x_{\text{B}}-O_{\text{B}}-y_{\text{B}}$ plane, with $O_{\text{B}} = [0,0]^{T}$ representing the local original point, a STARS equipped with $N$ MEs whose positions can be flexibly adjusted within the region $\mathcal{C}$ located in the $x_{\text{S}}-O_{\text{S}}-y_{\text{S}}$ plane, with $O_{\text{S}} = [0,0]^{T}$ representing the local original point, and $J$ single-antenna users. The local coordinate of the $m$-th FPA at the BS is fixed at $\mathbf{r}_m = \left[x_{m},y_{m}\right]^{T}$, while the local coordinate of the $n$-th ME at the STARS is restricted within the confined region by $\mathbf{u}_n = \left[x_{n},y_{n}\right]^{T}\in\mathcal{C}, \forall 1\leq n\leq N$. Let ${{\mathbf{U}} = \left[\mathbf{u}_1, \mathbf{u}_2, ..., \mathbf{u}_N\right]} \in\mathbb{R}^{2\times N}$ denote the STARS element position matrix (EPM). 
The reflection- and transmission-coefficient matrices of the ME-STARS are respectively given by 
\begin{equation}
\mathbf{\Theta}_{\text{r}} = \text{diag}\left(\sqrt{\beta_1^{\text{r}}}e^{j\theta_1^{\text{r}}},\sqrt{\beta_2^{\text{r}}}e^{j\theta_2^{\text{r}}}, ..., \sqrt{\beta_N^{\text{r}}}e^{j\theta_N^{\text{r}}}\right)
\end{equation}
\begin{equation}
\mathbf{\Theta}_{\text{t}} = \text{diag}\left(\sqrt{\beta_1^{{\text{t}}}}e^{j\theta_1^{{\text{t}}}},\sqrt{\beta_2^{{\text{t}}}}e^{j\theta_2^{{\text{t}}}}, ..., \sqrt{\beta_N^{{\text{t}}}}e^{j\theta_N^{{\text{t}}}}\right), 
\end{equation}
where $\beta_n^{\text{r}}$, $\beta_n^{{\text{t}}}\in \left[0,1\right]$ and   $\theta_n^{\text{r}}$, $\theta_n^{{\text{t}}}\in\left[0, 2\pi\right), \forall 1\leq n\leq N$ represent the amplitude and and phase-shift response of the $n$-th ME for the transmission and reflection, respectively. To ensure the energy conservation principle, the constraint $\beta_n^{\text{r}} + \beta_n^{{\text{t}}} = 1$ should be satisfied.  
The sets of users located in the STARS transmission and reflection spaces are denoted by $\mathcal{J}_r$ and $\mathcal{J}_t$, respectively. 
% \begin{figure}
% 	\centering
% 	\includegraphics[scale=0.12]{Fig/system_model1.eps}
% 	\caption{Illustration of the MA-enabled STARS aided downlink communications.}
% 	\label{fig:system_model} 
% \end{figure} 
\begin{figure}
        \centering
        \includegraphics[scale=0.14]{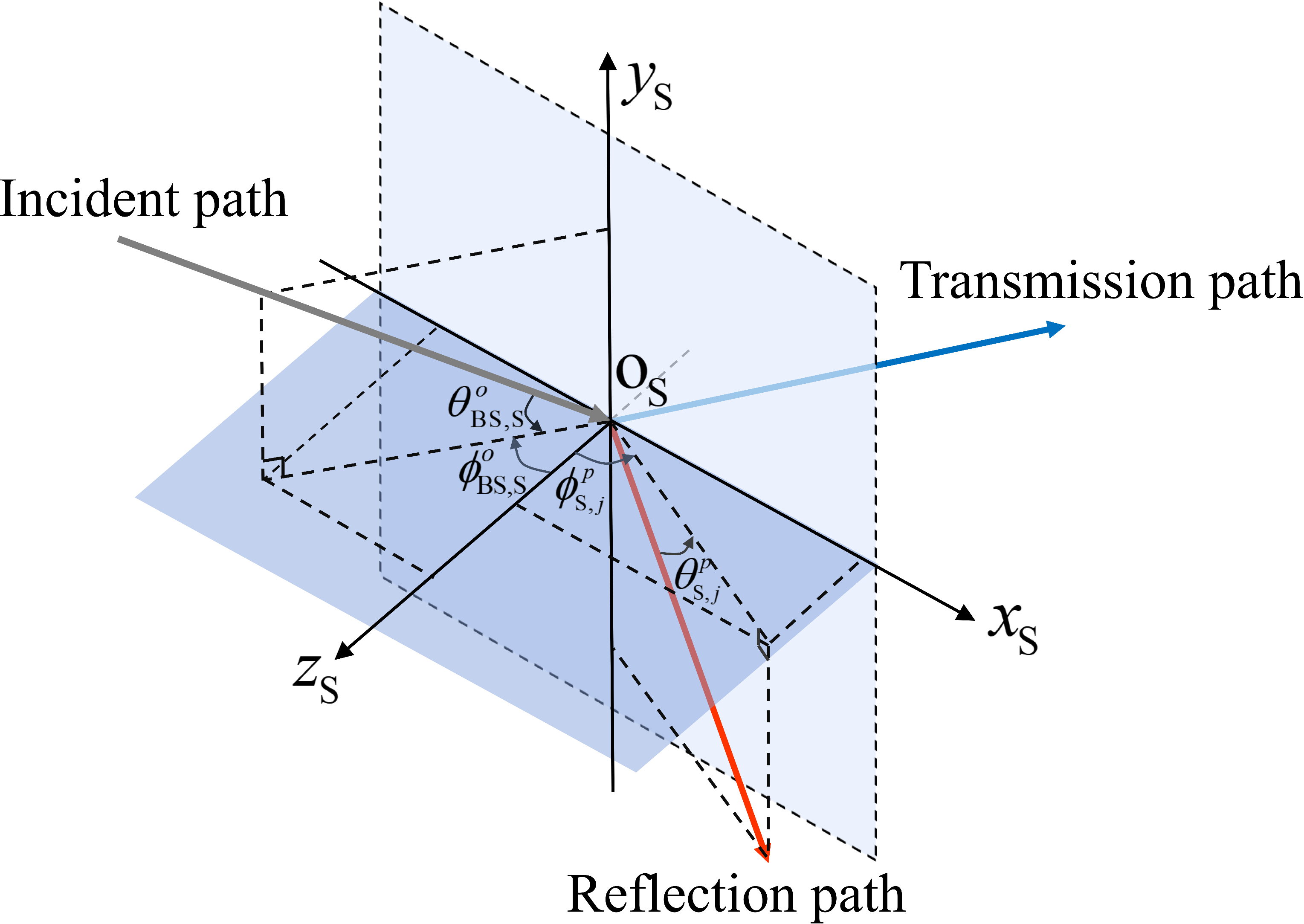}
        \caption{ME-STARS local coordinate system and the corresponding spatial angles.}
        \label{fig:STARS-angle}
    \label{fig:system_model_combined}
\end{figure}
\subsection{Channel Model}
The ME-STARS can not only reconfigure the BS-user communication environment by constructing the transmission/reflection paths, but also change the channel responses by flexibly adjusting the elements positions. In this work, we assume that the direct communication links between the BS and users are blocked by obstacles, and the BS-STARS-users cascased channels follow slow fading and we focus on one quasi-static fading block. Moreover, the far-field assumption is made for all the communication links, which is based on the setup that the signal propagation distance is much larger than the BS antenna aperture and the MEs moving region. Therefore, the plane-wave model can be used to form all the channel responses. 
Considering the geometric channel where there exists $L_{\text{BS}}$ and $L_{\text{S},j}$ paths of the BS-STARS and the STARS-user channels, respectively. Denote the elevation and azimuth angle-of-departure (AoD)/angle-of-arrival (AoA) of the $o$-th path between the BS and the ME-STARS as $\theta_{\text{BS,B}}^o$ ($\theta_{\text{BS,S}}^o$) $\in \left[-\pi/2, \pi/2\right]$ and $\phi_{\text{BS,B}}^o$ ($\phi_{\text{BS,S}}^o$) $\in \left[-\pi/2, \pi/2\right]$, respectively. The field-response vector (FRV) of the $m$-th antenna at the BS is given by~\cite{10318061}
\begin{equation}
    \mathbf{e}\left(\mathbf{r}_m\right) = \left[e^{j\frac{2\pi}{\lambda}\rho_{\text{B}}^{1}(\mathbf{r}_m)},e^{j\frac{2\pi}{\lambda}\rho_{\text{B}}^{2}(\mathbf{r}_m)}, ..., e^{j\frac{2\pi}{\lambda}\rho_{\text{B}}^{L_{\text{BS}}}(\mathbf{r}_m)}\right]^{T},
\end{equation}
where $\rho_{\text{B}}^{o}(\mathbf{r}_m) = x_m\cos{\theta_{\text{BS,B}}^o}\sin{\phi_{\text{BS,B}}^o}+y_m\sin{\theta_{\text{BS,B}}^o}$ is the signal propagation difference between the $m$-th FPA position $\mathbf{r}_m$ and the reference point $O_{\text{b}}$ for the $o$-th path. Then, the field response matrix (FRM) over all the FPAs at the BS can be represented by $\mathbf{E}=\left[\mathbf{e}\left(\mathbf{r}_1\right), \mathbf{e}\left(\mathbf{r}_2\right), ..., \mathbf{e}\left(\mathbf{r}_M\right)\right]\in\mathbb{C}^{L_{\text{BS}}\times M}$, which is a constant matrix as the antenna positions at the BS are fixed. 

Similarly, as shown in Fig.~\ref{fig:STARS-angle}, the FRV of the $n$-th ME at the STARS for the incident channel is given by
\begin{equation}
    \mathbf{f}_{\text{in}}(\mathbf{u}_n) = \left[e^{j\frac{2\pi}{\lambda}\rho_{\text{S,in}}^{1}(\mathbf{u}_n)},e^{j\frac{2\pi}{\lambda}\rho_{\text{S,in}}^{2}(\mathbf{u}_n)}, ..., e^{j\frac{2\pi}{\lambda}\rho_{\text{S,in}}^{L_{\text{BS}}}(\mathbf{u}_n)}\right]^{T},
\end{equation}
with $\rho_{\text{S,in}}^{o}(\mathbf{u}_n) = x_n\cos{\theta_{\text{BS,S}}^o}\sin{\phi_{\text{BS,S}}^o}+y_n\sin{\theta_{\text{BS,S}}^o}$ representing the signal propagation difference between the $n$-th ME position $\mathbf{u}_n$ and the reference point $O_{\text{s}}$ for the $o$-th path. By combining all the FRVs, we obtain the FRM at the ME-STARS for the incident channel as $\mathbf{F}_{\text{in}}\left(\mathbf{R}\right) = \left[\mathbf{f}_{\text{in}}(\mathbf{u}_1),\mathbf{f}_{\text{in}}(\mathbf{u}_2),...,\mathbf{f}_{\text{in}}(\mathbf{u}_N)\right]\in\mathbb{C}^{L_{\text{BS}}\times N}$. Denote the response from the BS reference point $O_{\text{B}}$ to the ME-STARS reference point $O_{\text{S}}$ as $\mathbf{\Sigma}_{\text{BS}} \in \mathbb{C}^{L_{\text{BS}}\times L_{\text{BS}}}$, then we obtain the BS-STARS channel response $\mathbf{H}\left({\mathbf{U}}\right)$ as follows:
\begin{equation}
    \mathbf{H}\left(\mathbf{U}\right) = \mathbf{F}_{\text{in}}^{{H}}\left(\mathbf{U}\right)\mathbf{\Sigma}_{\text{BS}}\mathbf{E}.
    \label{eq:BS-channel}    
\end{equation}

The FRV of the $n$-th ME at the STARS for the user $j\in\mathcal{J}_{\kappa}, \kappa\in\left\{t,r\right\}$ can be expressed as
\begin{equation}
    \mathbf{f}_{j}\left(\mathbf{u}_n\right) = \left[e^{j\frac{2\pi}{\lambda}\rho_{\text{S},j}^{1}(\mathbf{u}_n)},e^{j\frac{2\pi}{\lambda}\rho_{\text{S},j}^{2}(\mathbf{u}_n)}, ..., e^{j\frac{2\pi}{\lambda}\rho_{\text{S},j}^{L_{\text{S},j}}(\mathbf{u}_n)}\right]^{T},
\end{equation}
where $\rho_{\text{S},j}^p\left(\mathbf{u}_n\right) = x_n\cos{\theta_{\text{S},j}^p}\sin{\phi_{\text{S},j}^p}+y_n\sin{\theta_{\text{S},j}^p}$ with $\theta_{\text{S},j}^p$ and  $\phi_{\text{S},j}^p$ representing the AoDs for the $p$-th path between the STARS and the user $j$. We define the response from the ME-STARS reference point $O_{\text{s}}$ to the user $j$ as $\mathbf{\Sigma}_{\text{S},j}\in\mathbb{C}^{L_{\text{S},j}\times L_{\text{S},j}}$.
With $\mathbf{F}_{j}\left(\mathbf{U}\right) = \left[\mathbf{f}_{j}(\mathbf{u}_1),\mathbf{f}_{j}(\mathbf{u}_2),...,\mathbf{f}_{j}(\mathbf{u}_N)\right]\in\mathbb{C}^{L_{\text{S},j}\times N}$, we obtain the STARS-user channel response $\mathbf{g}_{\text{S},j}\left({\mathbf{U}}\right)$ as follows:
\begin{equation}
    \mathbf{g}_{\text{S},j}\left(\mathbf{U}\right) = \mathbf{1}_{L_{\text{S},j}}\mathbf{\Sigma}_{\text{S},j}\mathbf{F}_j\left(\mathbf{U}\right).
    \label{eq:SU-channel}
\end{equation}
\subsection{STARS Operating Protocols}
In this work, the three operating protocols proposed in~\cite{mu2021simultaneously} are also considered for the ME-STARS, namely ES, MS, and TS. For ES, each ME can reflect and transmit the incident signals simultaneously, which provides high degree of flexibility for the passive beamforming design. Accordingly, the feasible set of the transmission- and reflection-coefficients for ES is given by 
\begin{equation}
    \mathcal{F}^{\text{ES}} = \left\{\mathbf{\Theta} ^{\text{ES}}\mid \theta_{n}^{\kappa} \in [0,2\pi), \beta_{n}^{\kappa}\in\left[0, 1\right], \beta_{n}^{r} + \beta_{n}^{t} = 1\right\}.
\end{equation}
For MS, each ME works either in the transmission or the reflection mode. In other words, MS can be treated as the special case of ES, where the amplitude coefficients for the transmission and reflection are restricted to binary variables. Accordingly, the feasible set of the transmission- and reflection-coefficients for MS is given by 
\begin{equation}
    \mathcal{F}^{\text{MS}} = \left\{\mathbf{\Theta}^{\text{MS}} \mid  \theta_{n}^{\kappa} \in [0,2\pi), \beta_{n}^{\kappa}\in\left\{0, 1\right\}, \beta_{n}^{r} + \beta_{n}^{t} = 1\right\}.
\end{equation}
When the STARS employs ES or MS protocols, the received signal at user $j\in\mathcal{J}_\kappa$, $\kappa\in\left\{t,r\right\}$ is expressed as
\begin{align}
    y_{j} = & \mathbf{g}_{\text{S},j}\left({\mathbf{U}}\right)\mathbf{\Theta}_{\kappa}^{\text{ES}/\text{MS}}\mathbf{H}\left({\mathbf{U}}\right)\mathbf{w}_jx_j \notag\\
    & + \sum_{j'\neq j}^{J}\mathbf{g}_{\text{S},j}\left({\mathbf{U}}\right)\mathbf{\Theta}_{\kappa}^{\text{ES}/\text{MS}}\mathbf{H}\left({\mathbf{U}}\right)\mathbf{w}_{j'}x_{j'} + n_{j},
\end{align}
where $\mathbf{H}\left({\mathbf{U}}\right)\in\mathbb{C}^{N\times M}$ and $\mathbf{g}_{\text{S},j}\left({\mathbf{U}}\right)\in\mathbb{C}^{1 \times N}$
denote the BS-STARS and the STARS-user channels, respectively, $\mathbf{w}_j\in\mathbb{C}^{M\times 1}$ represents the BS beamforming vector for user $j$, $x_j$ is the signal transmitted from the BS to user $j$ with $\mathbb{E}\left[\left|x_j\right|^2\right]= 1$, and $n_j \sim \mathcal{CN}(0, \sigma_j^2)$ is the additive white Gaussian noise (AWGN) at user $j$. As such, the achievable communication rate of user $j$ is given by
\begin{align}
\label{eq:data-rate}
&R_j^{\text{ES}/\text{MS}} \nonumber\\
&= \log_2\left(1 + \frac{\left|\mathbf{g}_{\text{S},j}\left({\mathbf{U}}\right)\mathbf{\Theta}_{\kappa}^{\text{ES}/\text{MS}}\mathbf{H}\left({\mathbf{U}}\right)\mathbf{w}_j\right|^2}{\sum\limits_{\substack{j'\in \mathcal{J}\\ j'\neq j}}\left|\mathbf{g}_{\text{S},j}\left({\mathbf{U}}\right)\mathbf{\Theta}_{\kappa}^{\text{ES}/\text{MS}}\mathbf{H}\left({\mathbf{U}}\right)\mathbf{w}_{j'}\right|^2 + \sigma_j^2}\right).
\end{align}

Compared to ES and MS, TS exploits the time domain for covering the users in the transmission and reflection regions in a time-division manner. Since all the MAs are switched to the transmission or reflection mode simultaneously in each time slot, the feasible set of $\mathbf{\Theta}_{\kappa}$ is given by
\begin{equation}
    \mathcal{F}^{\text{TS}} = \left\{\mathbf{\Theta}^{\text{TS}} = \text{diag}\left(e^{j\theta_1^{\text{r}}}, ..., e^{j\theta_N^{\text{r}}}\right)\mid \theta_{n}^{\kappa} \in [0,2\pi)\right\}.
\end{equation}
For TS, let $\tau_{\kappa}, \kappa\in\left\{t,r\right\}$ denote the time allocation efficient for the transmission and reflection periods, which satisfies $\tau_{r}+\tau_{t}=1$. As such, the achievable communication rate of user $j\in\mathcal{J}_\kappa$, $\kappa\in\left\{t,r\right\}$ is given by
\begin{align}
\label{eq:TS-data-rate}
& R_j^{\text{TS}} \nonumber\\
&= \tau_{\kappa}\log_2\left(1 + \frac{\left|\mathbf{g}_{\text{S},j}\left({\mathbf{U}}\right)\mathbf{\Theta}_{\kappa}^{\text{TS}}\mathbf{H}\left({\mathbf{U}}\right)\mathbf{w}_j\right|^2}{\sum\limits_{\substack{j'\in \mathcal{J}_{\kappa}\\ j'\neq j}}\left|\mathbf{g}_{\text{S},j}\left({\mathbf{U}}\right)\mathbf{\Theta}_{\kappa}^{\text{TS}}\mathbf{H}\left({\mathbf{U}}\right)\mathbf{w}_{j'}\right|^2 + \sigma_j^2}\right).
\end{align}

\subsection{Problem Formulation}
In this work, we aim at maximizing the system WSR by jointly optimizing the active beamforming at the BS, as well as the MEs positions and the passive transmission/reflection beamforming at the STARS, while satisfying the movable region and the transmission/reflection coefficients feasibility constraints. 
\subsubsection{ES and MS}
When the ES or MS protocol is employed at the ME-STARS, the optimization problem can be formulated as follows:
\begin{subequations}
\label{eq:optimization_problem}
\begin{equation}
\label{eq:objective-function}
	(\text{P}1): \max_{\mathbf{U},\mathbf{W}, \mathbf{\Theta}_{\kappa}}\sum_{j=1}^{J}\omega_jR_j^{\text{X}},
\end{equation}
\begin{equation}
\label{eq:move-region}
    {\rm{s.t.}} \ \ \mathbf{u}_n\in\mathcal{C}, 1\leq n\leq N,
\end{equation}
\begin{equation}
\label{eq:ME-minimum-distance}
    \left\|\mathbf{u}_n - \mathbf{u}_{n'}\right\|_2\geq D_0, 1\leq n\neq n'\leq N,
\end{equation}
%\begin{equation}
%    \label{eq:data-rate-constraint}
%    R_j \geq R_j^{\text{min}}, j\in\mathcal{J}_\kappa,
%\end{equation}
\begin{equation}
    \label{eq:BS-transmit-power-constraint}
  {\rm{Tr}}\left(\mathbf{W}^{\text{H}}\mathbf{W}\right)\leq P^{\text{max}}, 
\end{equation}
\begin{equation}
    \label{eq:Theta-constraint}
    \mathbf{\Theta}_{\kappa}^{\text{X}}\in\mathcal{F}^{\text{X}}, \forall \kappa\in\left\{t,r\right\},
\end{equation}
\end{subequations}
where $\text{X}\in\left\{\text{ES, MS}\right\}$, $\omega_j$ is the weight representing the priority of user $j$, $\mathbf{W} = \left[\mathbf{w}_1,\mathbf{w}_2,...,\mathbf{w}_J\right]\in\mathbb{C}^{M\times J}$ is the active beamforming matrix for all users at the BS. Constraint \eqref{eq:move-region} ensures the MEs move within the feasible region $\mathcal{C}$. $D_0$ in constraint \eqref{eq:ME-minimum-distance} is the minimum distance among adjacent MEs to avoid antenna coupling. $P^{\text{max}}$ in constraint \eqref{eq:BS-transmit-power-constraint} is the maximum transmit power at the BS. \eqref{eq:Theta-constraint} restricts the feasible values of the transmission and reflection coefficients. 

\subsubsection{TS}: When the TS protocol is employed at the ME-STARS, due to the limited MEs moving speed in practice, we assume that the MEs positions are constant in the considered time period. Denote the active beamformers as $\mathbf{W}_t$ and $\mathbf{W}_r$ when the STARS work in transmission and reflection modes, respectively. Then, the optimization problem is formulated as
\begin{subequations}
\label{eq:optimization_problem_TS}
\begin{equation}
\label{eq:objective-function_TS}
	(\text{P}2): \max_{\mathbf{U},\mathbf{W}_{\kappa}, \mathbf{\Theta}_{\kappa},\tau_{\kappa}}\sum_{j=1}^{J}\omega_jR_j^{\text{TS}},
\end{equation}
\begin{equation}
    \eqref{eq:move-region}, \eqref{eq:ME-minimum-distance},
\end{equation}
\begin{equation}
    \label{eq:BS-transmit-power-constraint-TS}
  {\rm{Tr}}\left(\mathbf{W}_{\kappa}^{\text{H}}\mathbf{W}_{\kappa}\right)\leq P^{\text{max}}, \forall \kappa\in\left\{t,r\right\},
\end{equation}
\begin{equation}
    \label{eq:Theta-constraint_TS}
    \mathbf{\Theta}_{\kappa}^{\text{TS}}\in\mathcal{F}^{\text{TS}}, \forall \kappa\in\left\{t,r\right\},
\end{equation}
\begin{equation}
    \label{eq:time-constraint}
    0\leq \tau_{t}, \tau_r \leq 1, \tau_{t}+\tau_r=1,
\end{equation}
\end{subequations}
where \eqref{eq:time-constraint} imposes the time allocation constraint. 

Both (P1) and (P2) are intractable to solve as $\mathbf{U}$, $\mathbf{W}\slash \mathbf{W}_{\kappa}$, $\mathbf{\Theta}_{\kappa}$, $\tau_{\kappa}$ are highly coupled, and the objective functions~\eqref{eq:objective-function}\slash \eqref{eq:objective-function_TS} are highly non-convex with respect to (w.r.t.) the involved variables. In the following, we propose efficient iterative algorithms to obtain the suboptimal solutions by invoking the AO.

\section{Proposed Solution of Joint MEs Positions and Beamforming Optimization}
In this section, we first propose an iterative algorithm to solve the joint MEs positions and beamforming optimization problem for ES. Then the proposed algorithm is extended to solve the corresponding problems for MS and TS, respectively.
\subsection{Proposed Solution for ES}
\label{sec:ES}
\subsubsection{Optimization of $\mathbf{U}$ with given $\mathbf{W}$ and $\mathbf{\Theta}_{\kappa}$}
This subproblem is to optimize $\mathbf{U}$ in (P1) with given $\mathbf{W}$ and $\mathbf{\Theta}_{\kappa}$, which can be expressed as
\begin{subequations}
\begin{equation}
\label{eq:R-objective-function}
	(\text{P1-1}): \max_{\mathbf{U}}f(\mathbf{U}) = \sum_{j=1}^{J}\omega_jR_j^{\text{ES}},
\end{equation}
\begin{equation}
\eqref{eq:move-region}, \eqref{eq:ME-minimum-distance}.
\end{equation}
\end{subequations}
Since $R_j^{\text{ES}}$ defined in~\eqref{eq:data-rate} is highly non-convex w.r.t. $\mathbf{u}_n$, it is challenging to obtain the global optimal solution for problem (P1-1). Next, we propose the GDA framework ~\cite{GDA} for optimizing $\mathbf{U}$. 

As observed in~\eqref{eq:data-rate}, $\mathbf{U}$ is a variable of both $\mathbf{g}_{\text{S},j}(\mathbf{U})$ and $\mathbf{H}(\mathbf{U})$, which makes the gradient derivation much complicated. To facilitate the subsequent calculations, we first transform $R_j$ into a more tractable form w.r.t. $\mathbf{U}$. Specifically, define the transmission- and reflection-coefficient vectors of the ME-STARS as $\mathbf{q}_{\kappa}=\left[\sqrt{\beta_1^{{\kappa}}}e^{j\theta_1^{\kappa}},\sqrt{\beta_2^{\kappa}}e^{j\theta_2^{\kappa}}, ..., \sqrt{\beta_N^{\kappa}}e^{j\theta_N^{\kappa}}\right]^{H}$, $\forall \kappa\in\left\{t,r\right\}$. Then we have $\left|\mathbf{g}_{\text{S},j}\left({\mathbf{U}}\right)\mathbf{\Theta_{\kappa}}\mathbf{H}\left({\mathbf{U}}\right)\mathbf{w}_{j}\right|^2 = \left|\mathbf{q}_{\kappa}^H\mathbf{V}_{j}\left(\mathbf{U}\right)\mathbf{w}_{j}\right|^2$. $\mathbf{V}_{j}\left(\mathbf{U}\right)\in\mathbb{C}^{N\times M}$ is given by
\begin{align}
\label{eq:cas-channel}
    & \mathbf{V}_{j}\left(\mathbf{U}\right)  = \text{diag}\left(\mathbf{g}_{\text{S},j}\left(\mathbf{U}\right)\right)\mathbf{H}(\mathbf{U})\nonumber \\
    & = \text{diag}\left(\mathbf{\sigma}_{\text{S},j}\mathbf{F}_j\left(\mathbf{U}\right)\right)\mathbf{F}_{\text{in}}^{{H}}\left(\mathbf{U}\right)\mathbf{\Sigma}_{\text{BS}}\mathbf{E}
   \nonumber\\ 
   & =\left(\mathbf{\sigma}_{\text{S},j}\mathbf{F}_j(\mathbf{u}_1)\mathbf{\Sigma}_{\text{BS}}\mathbf{E}, ..., \mathbf{\sigma}_{\text{S},j}\mathbf{F}_j(\mathbf{u}_N)\mathbf{\Sigma}_{\text{BS}}\mathbf{E}\right)^T,
\end{align}
where $\mathbf{F}_j(\mathbf{u}_n)=\mathbf{f}_j(\mathbf{u}_n)\left(\mathbf{f}_{\text{in}}(\mathbf{u}_n)\right)^H\in\mathbb{C}^{L_{\text{S},j}\times L_{\text{BS}}}$ is only determined by the position of the $n$-th ME, which can be expanded as in~\eqref{eq:F_j_expansion}. 
\begin{figure*}
\begin{align}
    \label{eq:F_j_expansion}
    \mathbf{F}_j(\mathbf{u}_n) = \left[
    \begin{matrix}
    e^{j\frac{2\pi}{\lambda}(\rho_{\text{S},j}^1(\mathbf{u}_n)-\rho_{\text{S,in}}^1(\mathbf{u}_n))} & ... & e^{j\frac{2\pi}{\lambda}(\rho_{\text{S},j}^1(\mathbf{u}_n)-\rho_{\text{S,in}}^{L_{\text{BS}}}(\mathbf{u}_n))}\\
    \vdots & \ddots & \vdots\\
    e^{j\frac{2\pi}{\lambda}(\rho_{\text{S},j}^{L_{\text{S},j}}(\mathbf{u}_n)-\rho_{\text{S,in}}^1(\mathbf{u}_n))} & ... & e^{j\frac{2\pi}{\lambda}(\rho_{\text{S},j}^{L_{\text{S},j}}(\mathbf{u}_n)-\rho_{\text{S,in}}^{L_{\text{BS}}}(\mathbf{u}_n))}
    \end{matrix}
    \right].
\end{align}
\end{figure*}
Accordingly,  $\left|\mathbf{q}_{\kappa}^H\mathbf{V}_{j}\left(\mathbf{U}\right)\mathbf{w}_{j}\right|^2$ can be rewritten in an element-wise manner as follows:
\begin{align} 
\label{eq:element-wise}\left|\mathbf{q}_{\kappa}^H\mathbf{V}_{j}\left(\mathbf{u}_n\right)\mathbf{w}_{j}\right|^2 = \left|\mathbf{a}_{n,j}\mathbf{F}_j(\mathbf{u}_n)\mathbf{b}_j + {\sum_{n'\neq n}^N \mathbf{a}_{n',j}\mathbf{F}_j(\mathbf{u}_{n'})\mathbf{b}_j}\right|^2,
\end{align}
where $\mathbf{a}_{n,j} = \mathbf{q}_{\kappa}^H[n]\mathbf{\sigma}_{\text{S},j} \in\mathbb{C}^{1\times L_{\text{S},j}}$, $\mathbf{a}_{n',j} = \mathbf{q}_{\kappa}^H[n']\mathbf{\sigma}_{\text{S},j} \in\mathbb{C}^{1\times L_{\text{S},j}}$, and $\mathbf{b}_j = \mathbf{\Sigma}_{\text{BS}}\mathbf{E}\mathbf{w}_{j}\in\mathbb{C}^{L_{\text{BS}}\times 1}$ are constant variables irrelevant to $\mathbf{U}$. Similarly, $\sum_{j'\neq j}^{J}\left|\mathbf{q}_{\kappa}^H\mathbf{V}_{j}\left(\mathbf{U}\right)\mathbf{w}_{j'}\right|^2$ can be rewritten in the element-wise manner as in~\eqref{eq:denominator}, where $\mathbf{b}_j' = \mathbf{\Sigma}_{\text{BS}}\mathbf{E}\mathbf{w}_{j'}\in\mathbb{C}^{L_{\text{BS}}\times 1}$. 
\begin{figure*}
    \begin{equation}
    \label{eq:denominator}
        \sum_{j'\neq j}^{J}\left|\mathbf{q}_{\kappa}^H\mathbf{V}_{j}\left(\mathbf{u}_n\right)\mathbf{w}_{j'}\right|^2 = \sum_{j'\neq j}^J\left|\mathbf{a}_{n,j}\mathbf{F}_j(\mathbf{u}_n)\mathbf{b}_{j'}+{\sum_{n'\neq n}^N\mathbf{a}_{n',j}\mathbf{F}_j(\mathbf{u}_{n'})\mathbf{b}_{j'}}\right|^2.
    \end{equation}
        \hrulefill
\end{figure*}

Given that the GDA can not solve the constrained optimization problem directly, we first convert (P1-1) to an unconstrained one. To satisfy the confined region constraint~\eqref{eq:move-region}, we introduce the auxiliary variable $\tilde{\mathbf{U}}\in\mathbb{R}^{2\times N}$ that satisfies 
\begin{equation}
    \mathbf{U} = \frac{A}{2}\tanh \left(\tilde{\mathbf{U}}\right),
    \label{eq:tanh}
\end{equation}
where 
\begin{equation}
    \tanh(\mathbf{X})[a,b] = \frac{e^{\mathbf{X}[a,b]}-e^{-\mathbf{X}[a,b]}}{e^{\mathbf{X}[a,b]}+e^{-\mathbf{X}[a,b]}}\in\left(-1,1\right).
\end{equation}
The above operation is to project the variables $\tilde{\mathbf{u}}_n, \forall n$ defined in the real space onto the confined real space $\mathcal{C}$. As such, (P1-1) can be transformed to the following optimization problem:
\begin{subequations}
\begin{equation}
\label{eq:R-objective-function}
	(\text{P1-1-1}): \max_{\tilde{\mathbf{U}}}f\left(\tilde{\mathbf{U}}\right),
\end{equation}
\begin{align} g\left(\tilde{\mathbf{u}}_n,\tilde{\mathbf{u}}_{n'}\right)=\frac{2D_0}{A}- & ||\tanh (\tilde{\mathbf{u}}_n) - \tanh (\tilde{\mathbf{u}}_{n'})||_2\leq 0, \nonumber\\
    & \forall 1\leq n\neq n'\leq N.
    \label{eq:distance-constraint-convert}
\end{align}
\end{subequations}
For the remaining constraint~\eqref{eq:distance-constraint-convert}, we employ the penalty method~\cite{penalty-2} to convert it to a penalty term in the objective function when the inequality is not satisfied. Specifically, the penalty term can be given by $\eta\sum_{1\leq n< N}\sum_{n'=n+1}\max \left\{0, g\left(\tilde{\mathbf{u}}_n,\tilde{\mathbf{u}}_{n'}\right)\right\}$, where $\eta>0$ is the penalty factor which implies the weight for penalizing the objective function when~\eqref{eq:distance-constraint-convert} is not satisfied. As the $\max$ function is non-differential, we adopt the log-sum-exp (LSE) function to smooth the max function following $\max\{a,b\}\approx \rho \ln \left(e^{a/\rho}+e^{b/\rho}\right)$ with $\rho>0$ being the smoothing parameter, where the approximation precision is increased with smaller $\rho$. 
As such, we can transform (P1-1-1) to (P1-1-2) as follows:
\begin{align}
    (\text{P1-1-2}): & \max_{\tilde{\mathbf{U}}}  \ p\left(\tilde{\mathbf{U}}\right) = f\left(\tilde{\mathbf{U}}\right)  \nonumber\\
    & - \eta_1\sum_{1\leq n< N}\sum_{n'= n+1}\rho\ln\left(1+ e^{g\left(\tilde{\mathbf{u}}_n,\tilde{\mathbf{u}}_{n'}\right)/\rho}\right).
    \label{eq:P1-1-3}
\end{align}
Now, (P1-1-2) is an unconstrained differential problem, which can be solved by the GDA. However, whether we can get the optimal/suboptimal optimization result of (P1-1-1) through solving (P1-1-2) highly depends on the values of the penalty factor $\eta_1$ and the smoothing parameter $\rho$. (P1-1-1) and (P1-1-2) can be treated as the same if $\eta$ is sufficiently large and $\rho$ is small enough~\cite{penalty}. However, if we put $\eta$ too large at the beginning, the objective function of (P1-1-2) is dominated by the penalty term and the solution might be restricted to the space where the sum rate maximization is not fully considered. Therefore, we propose to initialize $\eta_1$ with a relatively small value and gradually increases $\eta_1$. Meanwhile, $\rho$ is gradually decreased to improve the accuracy of the smoothing of the $\max$ function. The iterations repeat until eventually satisfying the constraint~\eqref{eq:distance-constraint-convert}. To this end, the algorithm for solving (P1-1-1) consists of a nested loop. In the outer loop, $\eta_1$ and $\rho$ are updated as $\eta_1 = \omega_{\eta}\eta_1$ with $\omega_{\eta}>1$ and $\rho = \omega_{\rho}\rho$ with $0<\omega_{\rho}<1$ until constraint~\eqref{eq:distance-constraint-convert} is satisfied. In the inner loop, $\tilde{\mathbf{u}}_n$ is optimized via the GDA, which is elaborated in the following. 

In the $i$-th iteration of the GDA, $\tilde{\mathbf{U}}^{(i)}$ is updated by moving along the gradient direction, i.e.,
\begin{equation}
    \tilde{\mathbf{U}}^{(i+1)} = \tilde{\mathbf{U}}^{(i)} + \tau^{(i)} \nabla p\left(\tilde{\mathbf{U}}^{(i)}\right),
    \label{eq:gradient-ascent-update}
\end{equation}
where $\nabla p\left(\tilde{\mathbf{U}}^{(i)}\right)\in\mathbb{R}^{2\times N}$ and $\tau^{(i)}$ are the gradient vector of $p\left(\tilde{\mathbf{U}}\right)$ at the point $\tilde{\mathbf{U}}^{(i)}$ and the corresponding step size. The gradient vector $\nabla p\left(\tilde{\mathbf{U}}^{(i)}\right)$ is calculated as follows:
\begin{equation}
\label{eq:nabla-p}
    \nabla p\left(\tilde{\mathbf{U}}\right) = \nabla p\left({\mathbf{U}}\right)\circ\frac{A}{2}\left(1-\tanh^2(\tilde{\mathbf{U}})\right),
\end{equation}
where $p\left({\mathbf{u}}_n\right)$ is the function obtained by substituting~\eqref{eq:tanh} into \eqref{eq:P1-1-3}, and $\circ$ is the Hadamard multiplication. The $n$-th column of $\nabla p\left({\mathbf{U}}\right)$ is given by 
\begin{align}
    \frac{\partial p\left({\mathbf{U}}\right)}{\partial \mathbf{u}_n} = & \sum_{j=1}^{J}\omega_j\frac{\partial R_j^{\text{ES}}}{\partial \mathbf{u}_n}\nonumber\\
    &-\eta_1 \sum_{n'\neq n}\frac{e^{g(\mathbf{u}_n,\mathbf{u}_{n'})/\rho}}{1+ e^{g(\mathbf{u}_n,\mathbf{u}_{n'})/\rho}}\frac{\partial g(\mathbf{u}_n,\mathbf{u}_{n'})}{\partial \mathbf{u}_n},
    \label{eq:partial-p}
\end{align}
where $g(\mathbf{u}_n,\mathbf{u}_{n'}) = D_0-\left\|\mathbf{u}_n-\mathbf{u}_{n'}\right\|_2$. $\frac{\partial g(\mathbf{u}_n,\mathbf{u}_{n'})}{\partial \mathbf{u}_n}$ is given by
\begin{equation}
    \frac{\partial g(\mathbf{u}_n,\mathbf{u}_{n'})}{\partial \mathbf{u}_n} = -\frac{\mathbf{u}_n-\mathbf{u}_{n'}}{\sqrt{\left(\mathbf{u}_n-\mathbf{u}_{n'}\right)^T\left(\mathbf{u}_n-\mathbf{u}_{n'}\right)}}.
    \label{eq:partial-g}
\end{equation}
The detailed derivation of ${\partial R_j^{\text{ES}}}/{\partial \mathbf{u}_n}$ is given in Appendix~\ref{app:A}.

In addition, we employ the backtracking line search to determine the value of the step size $\tau^{(i)}$~\cite{boyd2004convex}. Specifically, starting with a relatively large value, $\tau^{(i)}$ is shrinked with $\tau^{(i)} = \omega_{\tau}\tau^{(i)}$ until the following Armijo condition is satisfied: 
%\begin{align}
    %p\left(\tilde{\mathbf{U}}^{(i+1)}\right) & \geq p\left(\tilde{\mathbf{U}}^{(i)}\right)\nonumber\\&+\delta\tau^{(i)}\text{Tr}\left(\left(\nabla p\left((\mathbf{U}^{(i)}\right)\right)^T\nabla p\left(\mathbf{U}^{(i)}\right)\right),
%    \label{eq:Armijo-condition}
%\end{align}
\begin{equation}
    p\left(\tilde{\mathbf{U}}^{(i+1)}\right) \geq p\left(\tilde{\mathbf{U}}^{(i)}\right) +\delta\tau^{(i)}\left\|\nabla p\left(\mathbf{U}^{(i)}\right)\right\|_{\text{F}}^2,
    \label{eq:Armijo-condition}
\end{equation}
where $\delta\in\left(0,1\right)$ is the given parameter that controls the increment range of the objective function. 

The overall GDA for solving (P1-1) is given in \textbf{Algorithm~\ref{alg:U-optimization}}, which consists of an outer loop and an inner loop. The MAs positions $\mathbf{U}$ are initialized within the confined region $\mathcal{C}$ and assigned to $\tilde{\mathbf{U}}$ in lines 1-2. In lines 3-15, the outer loop iterates with increased accuracy for eventually satisfying the MAs minimum distance constraints. In lines 5-13, the MAs positions are iteratively updated with the GDA until the increment of $p\left(\tilde{\textbf{U}}\right)$ is below the threshold $\epsilon_1$ or the number of inner iterations reaches the threshold $i_{\text{max}}$. Since $\delta>0$, $\tau^{(i)}>0$ and $\|\nabla p(\mathbf{U}^{(i)})\|^2\geq 0$, it is ensured that $p\left(\tilde{\textbf{U}}\right)$ is non-decreasing in each iteration of the inner loop. As such, the inner loop is guaranteed to converge given that $p\left(\tilde{\textbf{U}}\right)$ is upper bounded. Moreover, with sufficiently large $\eta$ and relatively small $\rho$, the condition $\left\|\tanh\left(\tilde{\mathbf{u}}_n^{(i)}\right) - \tanh\left(\tilde{\mathbf{u}}_{n'}^{(i)}\right)\right\|_2\geq D_0, 1\leq n\neq n'\leq N$ can be eventually satisfied for finding the optimal solution of $p\left(\tilde{\textbf{U}}\right)$, which guarantees the convergence of the outer loop. Therefore, the solution of \textbf{Algorithm~\ref{alg:U-optimization}} after convergence is the stationary point for solving the original problem (P1-1).

Moreover, the computational complexity of \textbf{Algorithm~\ref{alg:U-optimization}} is analyzed as follows. The main complexity for calculating the gradient $\nabla p\left(\tilde{\mathbf{U}}^{(i)}\right)$
relies on the computation of 
$\partial R_j/{\partial \mathbf{u}_n^{(i)}}$ in line 6. Specifically, the complexities for calculating $\mathbf{a}_{n,j}$, $\mathbf{b}_{j}$, $c_{n,j}$, $i_{n,j}^{j'}$, $\nabla \upsilon_j(\mathbf{u}_n)$, and $\nabla \gamma_j(\mathbf{u}_n)$ are given by $\mathcal{O}(L_{\text{S},j})$, $\mathcal{O}(ML_{\text{BS}})$, $\mathcal{O}(NL_{\text{BS}}L_{\text{S},j})$, $\mathcal{O}(NL_{\text{BS}}L_{\text{S},j})$, $\mathcal{O}(L_{\text{BS}}L_{\text{S},j})$, and $\mathcal{O}(JL_{\text{BS}}L_{\text{S},j})$, respectively. Moreover, the complexity of the backtracking line search in line 10 is given by $\mathcal{O}\left(I_{\text{bt}}MN\right)$, where $I_{\text{bt}}$ is the number of searching steps. Therefore, the overall complexity of  \textbf{Algorithm~\ref{alg:U-optimization}} is $\mathcal{O}\left(I_1^{\text{out}}I_1^{\text{in}}\left(N^2L_{\text{BS}}L_{\text{S},j}+I_{\text{bt}}MN\right)\right)$, where $I_1^{\text{out}}$ and $I_1^{\text{in}}$ denote the number of outer and inner iterations required for convergence, respectively. As can be observed, the computational complexity of \textbf{Algorithm~\ref{alg:U-optimization}} is polynomial in $N$, which makes the practical implementation feasible even when the STARS is deployed with a large number of MEs. 

% The complexity for solving the problem (P1-2) is $\mathcal{O}\left((3J+N)^{1.5}\right)$ if the interior point method is employed. As such, the overall computational complexity of \textbf{Algorithm~\ref{alg:U-optimization}} is $\mathcal{O}\left(NI_{\text{in}}\left((3J+N)^{1.5} + JNL_{\text{BS}}L_{\text{S},j}\right)\right)$, where $I_{\text{in}}$ is the number of iterations for the convergence of the SCA. As can be observed, the computational complexity is polynomial in $J$, $N$, $L_{\text{BS}}$ and $L_{\text{S},j}$.

\begin{algorithm}[tp]
\caption{Algorithm for Solving Problem (P1-1)}
\label{alg:U-optimization}
\LinesNumbered
\KwIn{$\{\mathbf{r}_m\}_{m=1}^M$, $\mathbf{\Sigma}_{\text{BS}}$, $\left\{\mathbf{\sigma}_{\text{S},j}\right\}_{j=1}^J$, $\left\{\theta_{\text{BS,B}}^o\right\}_{o=1}^{L_{\text{BS}}}$, $\{\phi_{\text{BS,B}}^o\}_{o=1}^{L_{\text{BS}}}$, $\{\theta_{\text{BS,S}}^o\}_{o=1}^{L_{\text{BS}}}$, $\{\phi_{\text{BS,S}}^o\}_{o=1}^{L_{\text{BS}}}$, $\{\theta_{\text{S},j}^p\}_{p=1}^{L_{\text{S},j}}$, $\{\phi_{\text{S},j}^p\}_{p=1}^{L_{\text{S},j}}$, $\mathcal{C}$, $D_0$, $\mathbf{W}$, $\mathbf{\Theta}$}

Initialize the EPM $\mathbf{U}$, the penalty factor $\eta_1$, the smooothing parameter $\rho$, and the initial step size $\overline{\tau};$\\
Set $\tilde{\mathbf{U}}^{(0)} = \mathbf{U}$;\\
\Repeat{ $\left\|\tanh\left(\tilde{\mathbf{u}}_n^{(i)}\right) - \tanh\left(\tilde{\mathbf{u}}_{n'}^{(i)}\right)\right\|_2\geq D_0, 1\leq n\neq n'\leq N$}{
    Set iteration index $i=0$ for inner loop;\\
    \Repeat{$p\left(\tilde{\mathbf{U}}^{(i+1)}\right) - p\left(\tilde{\mathbf{U}}^{(i)}\right)\leq \epsilon_1$ \textbf{or} $i\geq i_{\text{max}}$}{
    Calculate ${\partial R_j^{\text{ES}}}/{\partial \mathbf{u}_n^{(i)}}$ in~\eqref{eq:partial-R-j};\\
    Calculate ${\partial g(\mathbf{u}_n^{(i)},\mathbf{u}_{n'}^{(i)})}/{\partial \mathbf{u}_n^{(i)}}$ in~\eqref{eq:partial-g};\\
    Calculate ${\partial p\left(\mathbf{U}^{(i)}\right)}/{\partial \mathbf{u}_n^{(i)}}$ in~\eqref{eq:partial-p};\\
    Calculate  $\nabla p\left(\tilde{\mathbf{U}}^{(i)}\right)$ in~\eqref{eq:nabla-p};\\
    Update $\tau^{(i)} = \omega_{\tau}\tau^{(i)}$ until~\eqref{eq:Armijo-condition} is satisfied;\\
    Update $\tilde{\mathbf{U}}^{(i+1)}$ according to~\eqref{eq:gradient-ascent-update};\\
    $i = i+1$;\\
    }
    $\eta_1 = \omega_{\eta}\eta_1$, $\rho = \omega_{\rho}\rho$;
}
\textbf{Output} $\mathbf{U} = A/2\tanh \left(\tilde{\mathbf{U}}^{(i)}\right)$.
\end{algorithm}

%\begin{algorithm}[h]
% \caption{Algorithm for ME Positions Optimization}
% \label{alg:U-optimization}
% \LinesNumbered
% \KwIn{$\{\mathbf{r}_m\}_{m=1}^M$,$\mathbf{\Sigma}_{\text{BS}}$, $\left\{\mathbf{\sigma}_{\text{S},j}\right\}_{j=1}^J$, $\left\{\theta_{\text{BS,B}}^o\right\}_{o=1}^{L_{\text{BS}}}$, $\{\phi_{\text{BS,B}}^o\}_{o=1}^{L_{\text{BS}}}$, $\{\theta_{\text{BS,S}}^o\}_{o=1}^{L_{\text{BS}}}$, $\{\phi_{\text{BS,S}}^o\}_{o=1}^{L_{\text{BS}}}$, $\{\theta_{\text{S},j}^p\}_{p=1}^{L_{\text{S},j}}$, $\{\phi_{\text{S},j}^p\}_{p=1}^{L_{\text{S},j}}$, $\mathcal{C}$, ${D}_0$, $\mathbf{W}$, $\mathbf{\Theta}_{\kappa}$}
% \KwOut{$\mathbf{U}$}
% Initialize $\left\{\mathbf{u}_n\right\}_{n=1}^N = \left\{\mathbf{u}_n^{(0)}\right\}_{n=1}^N$;\\
% \For{$n = 1\rightarrow{N}$}{
%     Set iteration index $i = 0$;\\
%     \While{Increment of the WSR is above a threshold $\epsilon_1>0$}{
%         Solve problem (P1-2) to obtain $\mathbf{u}_n^{(i+1)}$ and update $\mathbf{u}_n = \mathbf{u}_n^{(i+1)}$;\\
%         $i = i+1$;
%     }  
% }
% \end{algorithm}

\subsubsection{Optimization of $\mathbf{W}$ with Given $\mathbf{\Theta}_{\kappa}$ and $\mathbf{U}$}
%\textbf{Optimization of $\mathbf{W}$: Closed-form with the convex form.}
%
%\textbf{Optimization of $\Theta$: Lagrangian method proposed by Zhaolin.}
%
%\textbf{Optimization of $\mathbf{\tilde{u}}$: Projected GDA (Two-timescale design for MAarra-enabled multiuser uplink communications), Cunhua Pan ``Two time-scale..." accelerated GDA}
The subproblem to optimize $\mathbf{W}$ is a typical WSR maximization problem in the mutli-user MISO system, which has been studied extensively in the existing literatures. One well-known solving approach is the weighted minimum mean square error (WMMSE) method~\cite{4712693,5756489}. Specifically, $\mathbf{W}$ can be effectively solved by iteratively optimizing the MSE weights $\left\{\varpi_j\right\}_{j=1}^J$, the scaling parameters $\left\{v_j\right\}_{j=1}^J$, and $\mathbf{W}$. According to~\cite{5756489}, $\left\{\varpi_j\right\}_{j=1}^J$ and $\left\{v_j\right\}_{j=1}^J$ can be given in the closed form as
\begin{subequations}
    \begin{equation}
	\varpi_j = \omega_j\left(1 + \frac{| \mathbf{q}_\kappa^H\mathbf{V}_j\left(\mathbf{U}\right) \mathbf{w}_j|^2}{\sum_{j' \neq j} |\mathbf{q}_\kappa^H\mathbf{V}_j\left(\mathbf{U}\right)\mathbf{w}_{j'}|^2 + \sigma_j^2}\right), \forall j,
    \end{equation}

    \begin{equation}
	v_j = \frac{\mathbf{q}_\kappa^H\mathbf{V}_j\left(\mathbf{U}\right)\mathbf{w}_j}{\sum_{j'=1}^{J} \left|\mathbf{q}_\kappa^H\mathbf{V}_j\left(\mathbf{U}\right)\mathbf{w}_{j'}\right|^2 + \sigma_j^2}, \forall j,
    \end{equation}
\end{subequations}
respectively. 

For the optimization of $\mathbf{W}$, the following convex optimization problem needs to be solved with given $\left\{\varpi_j\right\}_{j=1}^J$ and $\left\{v_j\right\}_{j=1}^J$:
\begin{subequations}
    \begin{equation}
        \text{(P1-2)}: \min_{\mathbf{W}}\sum_{j=1}^J\varpi_je_j,
    \end{equation}
    \begin{equation}
        {\rm{Tr}}\left(\mathbf{W}^{\text{H}}\mathbf{W}\right)\leq P^{\text{max}},
    \end{equation}
\end{subequations}
where $e_j = 1 - 2 \Re\{ v_j^{*} \mathbf{q}_\kappa^H\mathbf{V}_j\left(\mathbf{U}\right)\mathbf{w}_j \} + |v_j|^2\sum_{j'=1}^J \left|\mathbf{q}_\kappa^H\mathbf{V}_j\left(\mathbf{U}\right) \mathbf{w}_{j'}\right|^2 + |v_j|^2\sigma_j^2$. The problem is a typical convex quadratic programming (QP) problem, for which the optimal solution can be effectively obtained via standard convex problem solvers such as CVX~\cite{cvx}. If the interior point method is employed~\cite{boyd2004convex}, the computational complexity for solving (P1-2) is given by $\mathcal{O}\left(JM^{3.5}\right)$. Thus, assuming that $I_2$ iterations are required for the convergence of iteratively optimizing $\left\{\varpi_j\right\}_{j=1}^J$, $\left\{v_j\right\}_{j=1}^J$, and $\mathbf{W}$, the overall complexity is given by $\mathcal{O}\left(I_2JM^{3.5}\right)$.

\subsubsection{Optimization of $\mathbf{\Theta}_{\kappa}^{\text{ES}}$ with Given $\mathbf{W}$ and $\mathbf{U}$}
For the subproblem of optimizing $\mathbf{\Theta}_{\kappa}^{\text{ES}}$, existing works have shown that SCA is an effective approach to find the local-optimal solution~\cite{8982186,mu2021simultaneously}. Define $\mathbf{Q}_{\kappa} = \mathbf{q}_{\kappa}\mathbf{q}_{\kappa}^H$, which satisfies $\mathbf{Q}_{\kappa}\succeq 0$, $\text{Rank}\left(\mathbf{Q}_{\kappa}\right)=1$ and $\text{Diag}\left(\mathbf{Q}_{\kappa}\right) = \boldsymbol{\beta}^{\kappa}$, where $\boldsymbol{\beta}^{\kappa} \triangleq \left[\beta^{\kappa}_1, ..., \beta^{\kappa}_N\right]$. Further define $\mathbf{C}_j= \left(\mathbf{V}_j(\mathbf{U})\mathbf{w}_j\right)\left(\mathbf{V}_j(\mathbf{U})\mathbf{w}_j\right)^H$ and $\mathbf{C}_j^{j'}= \left(\mathbf{V}_j(\mathbf{U})\mathbf{w}_{j'}\right)\left(\mathbf{V}_j(\mathbf{U})\mathbf{w}_{j'}\right)^H$. By invoking the slack variables $A_{j}$ and $B_{j}$, the $\mathbf{\Theta}_{\kappa}$ optimization subproblem can be formulated as
\begin{subequations}
	\label{eq:Theta-optimization}
	\begin{equation}
		\label{eq:Theta-objective-function}
		\text{(P1-3)}: \max_{\mathbf{Q}_{\kappa}, \boldsymbol{\beta}^{\kappa}, \left\{A_j\right\}, \left\{B_{j}\right\}, \left\{R_j\right\}}\sum_{j=1}^{J}\omega_jR_j^{\text{ES}},
	\end{equation}
	\begin{equation}
		\label{eq:Theta-C1}
		{\rm{s.t.}} \ \ \log_2\left(1+\frac{1}{A_jB_{j}}\right)\geq R_j^{\text{ES}}, \forall j,
	\end{equation}
	\begin{equation}
		\label{eq:Theta-C2}
		\frac{1}{A_j}\leq \text{Tr}\left(\mathbf{Q}_{\kappa}\mathbf{C}_j\right), \forall j,
	\end{equation}
	\begin{equation}
		\label{eq:Theta-C3}
		B_{n,j}\geq \sum_{j'\neq j}^J\text{Tr}\left(\mathbf{Q}_{\kappa}\mathbf{C}_j^{j'}\right)+\sigma_j^2, \forall j,
	\end{equation}
	\begin{equation}
		\label{eq:Theta-C4}
		\mathbf{Q}_{\kappa} \succeq 0, \text{Rank}(\mathbf{Q}_{\kappa}) = 1, \text{Diag}(\mathbf{Q}_{\kappa}) = \boldsymbol{\beta}^{\kappa}, \forall \kappa,
	\end{equation}
        \begin{equation}
        \label{eq:amplitude-constraint}
        \beta_{n}^{t}, \beta_{n}^{r}\in\left[0, 1\right], \beta_{n}^{t} + \beta_{n}^{r} = 1, \forall n.
        \end{equation}
\end{subequations}
\begin{algorithm}[tp]
	\caption{Algorithm for Solving Problem (P1-3)}
	\label{alg:theta-optimization-ES}
	\LinesNumbered
	\KwIn{$\{\mathbf{r}_m\}_{m=1}^M$, $\{\mathbf{u}_n\}_{n=1}^N$, $\mathbf{\Sigma}_{\text{BS}}$, $\left\{\mathbf{\sigma}_{\text{S},j}\right\}_{j=1}^J$, $\left\{\theta_{\text{BS,B}}^o\right\}_{o=1}^{L_{\text{BS}}}$, $\{\phi_{\text{BS,B}}^o\}_{o=1}^{L_{\text{BS}}}$, $\{\theta_{\text{BS,S}}^o\}_{o=1}^{L_{\text{BS}}}$, $\{\phi_{\text{BS,S}}^o\}_{o=1}^{L_{\text{BS}}}$, $\{\theta_{\text{S},j}^p\}_{p=1}^{L_{\text{S},j}}$, $\{\phi_{\text{S},j}^p\}_{p=1}^{L_{\text{S},j}}$, $\mathbf{W}$}
	\KwOut{$\mathbf{Q}_{\kappa}, \kappa\in\left\{t,r\right\}$}
	Initialize $\mathbf{Q}_{\kappa}^{(0)}$ and $\alpha$;\\
	\Repeat{$\max\left\{\text{Tr}\left( \mathbf{Q}_{\kappa} \right) - \|\mathbf{Q}_{\kappa}\|_2\right\} < \epsilon_3$}{
		Set iteration index $i = 0$ for inner loop;\\
		\Repeat{Increment of the WSR is below $\epsilon_2$ \textbf{or} $i\geq i_{\text{max}}$}{
			Solve the relaxed problem of~\eqref{eq:Theta-optimization} to obtain $\mathbf{Q}_{\kappa}^{(i)}$ and update $\mathbf{Q}_{\kappa} = \mathbf{Q}_{\kappa}^{(i)}$;\\
			$i = i+1$;
		}
		Update $\mathbf{Q}_{\kappa}^{(0)} = \mathbf{Q}_{\kappa}^{(i)}$ and $\eta_2 = \omega_{\eta}\eta_2$.
	}
    \label{alg:Theta-optimization-ES}
\end{algorithm}
The main challenge for solving the above optimization problem relies on the non-convexity of~\eqref{eq:Theta-C1} and the rank-one constraint in~\eqref{eq:Theta-C4}. For~\eqref{eq:Theta-C1}, we can apply the first-order Taylor expansion to transform it into a linear approximation as follows.
\begin{equation}
	\begin{aligned}
		\label{eq:ES-SCA}
		&\log_2\left(1+\frac{1}{A_{j}^i B_{j}^i}\right) - \frac{(A_{j} - A_{j}^i)}{(\ln 2)A_{j}^i\left(1 + A_{j}^iB_{j}^i\right)} \\
		& - \frac{(B_{j}-B_{j}^i)}{(\ln 2) B_{j}^i\left(1+A_{j}^iB_{j}^i\right)}  \geq R_j, \forall j,
	\end{aligned}
\end{equation}

For the rank-one constraint, it is first equivalently transformed into the following equality constraint~\cite{Globecom} 
\begin{equation}
\label{eq:rank-one-equality}
    \text{Tr}\left( \mathbf{Q}_{\kappa} \right) - \|\mathbf{Q}_{\kappa}\|_2 = 0, \forall \kappa \in\left\{t,r\right\}.
\end{equation}
Again, the penalty method can be utilized to treat~\eqref{eq:rank-one-equality} as a penalty term added to the objective function. As such, ~\eqref{eq:Theta-objective-function} is rewritten as 
\begin{equation}
	\label{eq:ES-reformulated-function}
    \max_{\mathbf{Q}_{\kappa},\boldsymbol{\beta}^{\kappa},\left\{A_j\right\}, \left\{B_{j}\right\}, \left\{R_j\right\}}\sum_{j=1}^{J}\omega_jR_j - \eta_2\sum_{\kappa\in\left\{r,t \right\}}\left(\text{Tr}\left( \mathbf{Q}_{\kappa} \right) - \|\mathbf{Q}_{\kappa}\|_2\right),
\end{equation}
where $\eta_2$ is the penalty factor for violating the rank-one constraint. 
Since the penalty term is non-convex and in the form of difference of convex (DC) functions, we can adopt the first-order Taylor expansion to construct the upper-bound surrogate function in the $i$-th iteration of the SCA method as follows:
\begin{align} \text{Tr}\left(\mathbf{Q}_{\kappa}\right) - \left\|\mathbf{Q}_{\kappa}\right\|_2\leq \text{Tr}\left(\mathbf{Q}_{\kappa}\right) - \left[\left\|\mathbf{Q}_{\kappa}\right\|_2\right]^i_{\text{{lb}}},    
\end{align}
where $\left[\left\|\mathbf{Q}_{\kappa}\right\|_2\right]^i_{\text{{lb}}} \triangleq \left\|\mathbf{Q}_{\kappa}^{(i)}\right\|_2 + \text{Tr} \left[ \bar{\mathbf{x}}^i \left(\bar{\mathbf{x}}^i\right)^H (\mathbf{Q}_{\kappa} - \mathbf{Q}_{\kappa}^{(i)}) \right]$ and $\bar{\mathbf{x}}^i$ denotes the eigenvector w.r.t. the largest eigenvalue of $\mathbf{Q}_{\kappa}^{(i)}$. To this end, the original problem (P1-3) can be iteratively solved by applying CVX upon the relaxed QP~\cite{cvx}. As shown in \textbf{Algorithm~\ref{alg:theta-optimization-ES}}, the proposed solution consists of the outer loop for iteratively updating $\eta_2$, and the inner loop for iteratively solving the relaxed version of (P1-3). The main computational complexity of \textbf{Algorithm~\ref{alg:theta-optimization-ES}} relies on applying CVX to solve the QP problem. If the interior point method is employed, the complexity is given by $\mathcal{O}\left(N^{3.5}\right)$~\cite{local-convergence}. Then, the overall complexity is $\mathcal{O}\left(I_2^{\text{out}}I_2^{\text{in}}N^{3.5}\right)$, where $I_2^{\text{out}}$ and $I_2^{\text{in}}$ are the number of outer and inner iterations, respectively.

\subsubsection{Overall AO Algorithm Design}
With the solutions obtained for problem (P1-1), (P1-2) and (P1-3), we now finalize the overall algorithm design for solving the original problem (P1), which is given in details in \textbf{Algorithm~3}. Specifically, the proposed algorithms for solving problem (P1-1), (P1-2), and (P1-3) are iteratively executed until the increment of the WSR is below the predefined convergence threshold $\epsilon_2$. Since the WSR is non-decreasing during the process in alternatively optimizing $\mathbf{U}, \mathbf{W}$, and $\mathbf{\Theta}_{r}$, $\mathbf{\Theta}_{t}$, the iterations in \textbf{Algorithm~\ref{alg:ES-algorithm}} is limited due to the finite WSR. In other words, \textbf{Algorithm~\ref{alg:ES-algorithm}} is guaranteed to converge to a solution of (P1) that is at least locally optimal. 

\begin{algorithm}[tp]
	\caption{Alternating Algorithm for Solving Problem (P1)}
        \label{alg:ES-algorithm}
	\LinesNumbered
	\KwIn{$\{\mathbf{r}_m\}_{m=1}^M$, $\mathbf{\Sigma}_{\text{BS}}$, $\left\{\mathbf{\sigma}_{\text{S},j}\right\}_{j=1}^J$, $\left\{\theta_{\text{BS,B}}^o\right\}_{o=1}^{L_{\text{BS}}}$, $\{\phi_{\text{BS,B}}^o\}_{o=1}^{L_{\text{BS}}}$, $\{\theta_{\text{BS,S}}^o\}_{o=1}^{L_{\text{BS}}}$, $\{\phi_{\text{BS,S}}^o\}_{o=1}^{L_{\text{BS}}}$, $\{\theta_{\text{S},j}^p\}_{p=1}^{L_{\text{S},j}}$, $\{\phi_{\text{S},j}^p\}_{p=1}^{L_{\text{S},j}}$, $\mathcal{C}$, $D_0$}
	Initialize $\mathbf{W}$, $\mathbf{{\Theta}}_t$, $\mathbf{{\Theta}}_r$\\
    \Repeat{Increment of the WSR is below $\epsilon_2$}{
    Given $\mathbf{W}$, $\mathbf{{\Theta}}_t$, $\mathbf{{\Theta}}_r$, solve problem (P1-1) to update $\mathbf{U}$;\\
    Given $\mathbf{U}$, $\mathbf{{\Theta}}_t$, $\mathbf{{\Theta}}_r$, solve problem (P1-2) to update $\mathbf{W}$;\\
    Given $\mathbf{U}$, $\mathbf{W}$, solve problem (P1-3) to update $\mathbf{{\Theta}}_t$, $\mathbf{{\Theta}}_r$;
    }
    \KwOut{$\mathbf{U}, \mathbf{W}, \mathbf{\Theta}_{t}, \mathbf{\Theta}_{r}$}
\end{algorithm}
\subsection{Extended Solutions for MS and TS}
MS can be treated as a special case of ES, where the amplitude coefficients of each ME are restricted to binary variables, i.e. $\beta_{n}^{\kappa}\in\left\{0,1\right\}, \forall \kappa \in\left\{t,r\right\}$. Therefore, the optimization of $\mathbf{\Theta}_{\kappa}$ needs to be re-designed by considering the non-convex integer constraints, while $\mathbf{U}$ and $\mathbf{W}$ can be optimized following the similar manner as previously described. According to~\cite{mu2021simultaneously}, by converting the binary constraints to the continuous equality ones, i.e., $\beta_{n}^{\kappa} - (\beta_{n}^{\kappa})^2=0, \forall \kappa \in\left\{t,r\right\}$, the constraints can be added to the objective function~\eqref{eq:ES-reformulated-function} as another penalty term. As the introduced penalty term is non-convex, the first-order Taylor expansion can be adopted to obtain the convex upper-bound. Subsequently, the relaxed problem can be efficiently solved via CVX~\cite{cvx}. The structure of the algorithm for solving $\mathbf{\Theta}_{\kappa}^{\text{MS}}$ is similar to ~\textbf{Algorithm~\ref{alg:theta-optimization-ES}}, except from the extra penalty term. The details are omitted here for brevity. 

For TS, $\mathbf{U}$, $\mathbf{W}_{\kappa}$ and $\mathbf{\Theta}_{\kappa}$ should be jointly optimized with the time allocation coefficients $\tau_{\kappa}, \kappa\in\left\{t,r\right\}$. Again, applying the AO method, the optimization of $\mathbf{U}$, $\mathbf{W}_t$, $\mathbf{W}_r$ can follow the similar approach as previously described for ES. Different from ES\slash MS, as the transmission and reflection modes are activiated successively in orthogonal time slots, the optimization of $\mathbf{\Theta}_t$ and $\mathbf{\Theta}_r$ can be decoupled for TS, which yields the following subproblem:
\begin{subequations}
	\label{eq:Theta-optimization-TS}
	\begin{equation}
		\label{eq:Theta-objective-function-TS}
		\max_{\mathbf{q}_{t}/\mathbf{q}_{r}}\sum_{j=1}^{J} \omega_j R_j^{\text{TS}},
	\end{equation}
	\begin{equation}
	\label{eq:Theta-C1-TS}
	{\rm{s.t.}} \ \ \left|\left[\mathbf{q}_t\right]_{n}\right|=1 \ \text{or} \ \left|\left[\mathbf{q}_r\right]_{n}\right|=1, \forall n.
\end{equation}
\end{subequations}
The solution of the above optimization problem can be obtained via the gradient descent or the SCA methods~\cite{8982186}. Moreover, the time coefficients optimization is a typical resource allocation problem. Due to the page limit, the solving approaches for $\mathbf{\Theta}_{\kappa}$ 
and $\tau_{\kappa}$ are omitted here. We refer the reader to~\cite{mu2021simultaneously} for a thorough explanation of the details. The overall alternating algorithm for TS is summarized in \textbf{Algorithm~\ref{alg:TS-algorithm}}.
\begin{algorithm}[tp]
	\caption{Alternating Algorithm for Solving Problem (P2)}
	\label{alg:TS-algorithm}
	\LinesNumbered
	\KwIn{$\{\mathbf{r}_m\}_{m=1}^M$, $\mathbf{\Sigma}_{\text{BS}}$, $\left\{\mathbf{\sigma}_{\text{S},j}\right\}_{j=1}^J$, $\left\{\theta_{\text{BS,B}}^o\right\}_{o=1}^{L_{\text{BS}}}$, $\{\phi_{\text{BS,B}}^o\}_{o=1}^{L_{\text{BS}}}$, $\{\theta_{\text{BS,S}}^o\}_{o=1}^{L_{\text{BS}}}$, $\{\phi_{\text{BS,S}}^o\}_{o=1}^{L_{\text{BS}}}$, $\{\theta_{\text{S},j}^p\}_{p=1}^{L_{\text{S},j}}$, $\{\phi_{\text{S},j}^p\}_{p=1}^{L_{\text{S},j}}$, $\mathcal{C}$, $D_0$}
	Initialize $\mathbf{W}_t$, $\mathbf{W}_r$, $\mathbf{{\Theta}}_t$, $\mathbf{{\Theta}}_r$, $\tau_t$, $\tau_r$\\
    \Repeat{Increment of the WSR is below $\epsilon_2$}{
    Given $\mathbf{W}_t$, $\mathbf{W}_r$, $\mathbf{{\Theta}}_t$, $\mathbf{{\Theta}}_r$, $\tau_t$, $\tau_r$, update $\mathbf{U}$ via \textbf{Algorithm~\ref{alg:U-optimization}} by substituting $R_j^{\text{ES}}$ with $R_j^{\text{TS}}$;\\
    Given $\mathbf{U}$, $\mathbf{{\Theta}}_t\slash \mathbf{{\Theta}}_r$, update $\mathbf{W}_t\slash \mathbf{W}_r$ based on WMMSE method;\\
    Given $\mathbf{U}$, $\mathbf{W}_t\slash \mathbf{W}_r$, update $\mathbf{{\Theta}}_t\slash \mathbf{{\Theta}}_r$ by solving~\eqref{eq:Theta-optimization-TS};\\
     Solve the time coefficients optimization subproblem;
    }
    \KwOut{$\mathbf{U}$, $\mathbf{W}_t$, $\mathbf{W}_r$, $\mathbf{{\Theta}}_t$, $\mathbf{{\Theta}}_r$, $\tau_t$, $\tau_r$}
\end{algorithm}

%\begin{remark}
%    In the ES and MS modes, the optimal $\mathbf{U}$ is expected to well match the beam pattern for serving all the users. Whereas in the TS mode, due to the different timescales for the optimization of the EPM $\mathbf{U}$ and the beamformers $\mathbf{W}$, $\mathbf{\Theta}_{\kappa}$, the optimization of $\mathbf{U}$ should balance the tradeoff between the independent beam patterns for different sets of users in the transmission and reflection regions. 
%\end{remark}

\section{Numerical Results}
In this section, numerical results are provided to validate the effectiveness of the proposed ME-STARS aided downlink multiuser MISO communications system.
\subsection{Simulation Setup}
\begin{figure}
	\centering
	\includegraphics[scale=0.6]{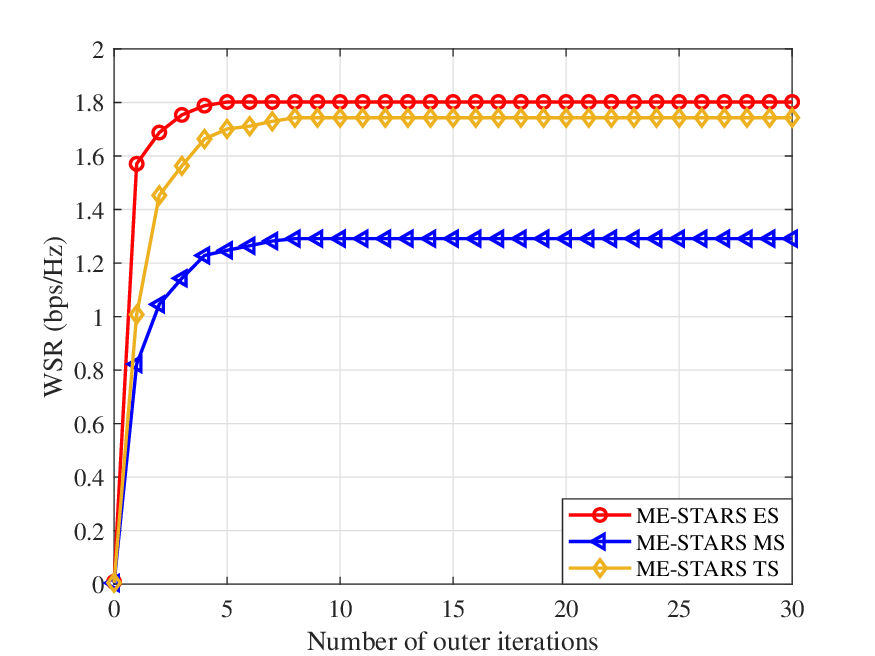}
	\caption{{Convergence performance of the proposed AO algorithms.}}
	\label{fig:AO-convergence} 
\end{figure}

Referring to the $x_{\text{S}}$-$y_{\text{S}}$-$z_{\text{S}}$ coordinate in Fig.~\ref{fig:STARS-angle}, the location of the BS is set at
$(-10,-5,10)$~meters, while the users are randomly distributed within a square region centered at $(0,-10,0)$~meters with the edge of $40$~meters in the $x_{\text{S}}-z_{\text{S}}$ plane. Considering the geometric channel model,
we assume that the number of paths is the same for all the considered channels, i.e., $L_{\text{BS}}=L_{\text{S},j}=L, \forall j$. $\mathbf{\Sigma}_{\text{BS}}$ is a diagonal matrix with each diagonal element following the circularly symmetric complex Gaussian (CSCG) distribution $\mathcal{CN}\left(0,\beta_{\text{BS}}/L\right)$, where $\beta_{\text{BS}}=\beta_0d_{\text{BS}}^{-\alpha_{0}}$ is the expected BS-STARS channel power gain, $\beta_0$ denotes the channel power gain at the reference distance of $1$~m, $d_{\text{BS}}$ represents the distance between the BS and the ME-STARS, and $\alpha_0$ is the path-loss exponent. Similarly, each element of $\mathbf{\sigma}_{\text{S},j}$ follows the CSCG distribution $\mathcal{CN}\left(0,\beta_{\text{S},j}/L\right)$, where $\beta_{\text{S},j}=\beta_0d_{\text{S},j}^{-\alpha_{0}}$, and $d_{\text{S},j}$ denotes the distance between the ME-STARS and the user $j$. The path angles $\theta
_{\text{BS,B}}^o$, $\theta_{\text{BS,S}}^o$, $\phi
_{\text{BS,B}}^o$, $\phi
_{\text{BS,S}}^o$, $\theta
_{\text{S},j}^p$, $\phi
_{\text{S},j}^p$, $\forall j, o, p$ are generated randomly within the range $\left[-\pi/2,\pi/2\right]$. To guarantee the users fairness, the weights $\omega_j, \forall j$ are chosen inversely proportional to the expected STARS-user channel power gain $\beta_{\text{S},j}$, and then normalized by $\sum \omega_j=1$. The adopted settings of simulation parameters are provided in Table~\ref{table:simulation-setup}, unless otherwise stated. Each point in the simulation figures (Figs.~4 - 8) is obtained by generating $10^3$ user distributions and channel realizations, except from the convergence demonstration in Fig.~\ref{fig:AO-convergence}, where the results are obtained over one random channel realization.
\begin{table}[h]
\caption{Simulation parameters}
\centering
\label{table:simulation-setup}
\begin{tabular}{|p{1cm}|p{5.4cm}|p{1cm}|}
\hline
Parameter & Description & Value\\
\hline
$f$ & Carrier frequency & $3$~GHz \\ \hline
$M$ & Number of FPEs at the BS  & $8$\\ \hline
$N$ & Number of MEs at the STARS  & $8$ \\ \hline
$J$ & Number of users  & $4$\\ \hline
$L$ & Number of paths for each channel  & $2$\\ \hline
$\beta_0$ & Channel power gain at $1$ meter  & $-30$~dB\\ \hline
$\alpha_0$ & Path-loss exponent  & $2.2$\\ \hline
$P^{\text{max}}$ & BS maximum transmit power & $30$~dBm\\ \hline
$\sigma_j^2$ & User noise power  & $-90$~dBm\\ \hline
$A$ & Length of the sides of moving region $\mathcal{C}$ & $2.5\lambda$\\ \hline
$\overline{\tau}$ & Initial gradient-ascent step size in Algorithm 1 & $10$ \\ \hline
$\omega_{\tau}$ & Step size shringking parameter in Algorithm 1 & $0.5$ \\ \hline
$\epsilon_1$, $\epsilon_2$ & Increment threshold in {Algorithm 1-4} & $10^{-6}$ \\ \hline
$i_{\text{max}}$ & {Max. number of iterations in {Algorithm 1 \& 2}} & $100$ \\ \hline
$\epsilon_3$ & Accuracy for constraint in Algorithm 2 & $10^{-7}$ \\ \hline
${\eta}_1$, ${\eta}_2$ & Initial penalty factors in Algorithm 1 \& 2 & $10^{-4}$ \\ \hline
${\rho}$ & Initial smoothing parameter in Algorithm 1 & $1$ \\ \hline
$\omega_{\eta}$ & Scaling factor in Algorithm 1 \& 2 & $10$ \\ \hline
$\omega_{\rho}$ & Scaling factor in Algorithm 1 & $0.1$ \\ \hline
\end{tabular}
\end{table}

To demonstrate the effectiveness of the ME-STARS aided communications, we choose the following benchmarks for comparison:
\begin{itemize}
	\item \textbf{FPE-STARS (STARS with fixed-position elements)}: In this case, we assume that the STARS is equipped with $N$ FPEs spaced by $\lambda/2$. The WSR maximization problem is addressed by iteratively solving the active beamforming and the passive transmission/reflection beamforming subproblems via the WMMSE and the SCA methods, respectively, that have been proposed in Section III.
    \item \textbf{ME-RIS (RIS with movable elements)}: In this case, one conventional reflecting-only RIS and one transmitting-only RIS are deployed at the same location as the ME-STARS to provide full-space coverage. For fair comparison, it is assumed that each RIS is deployed with $\frac{N}{2}$ MEs. This baseline can be treated as a special case of the MS mode, where half of the MEs work in the transmission mode and the other half work in the reflection mode. 
 %	\item \textbf{Random Phase (RP)}: The reflecting/transmitting parameters of STARS $\mathbf{\Theta}_{\kappa}$ is randomly initialized and not involved in optimization. The active precoding matrices $\mathbf{W}$ and Movable Positions $\mathbf{U}$ together solve the WSR maximization problem.
\end{itemize}

In Fig.~\ref{fig:AO-convergence}, we demonstrate the convergence performance of the proposed AO algorithms for the joint MEs positions and beamforming optimization problems for ES, MS, and TS. It is shown that the WSR obtained by all the three algorithms increases rapidly at the beginning of the iterations. For example, the AO algorithm converge within around {$5$} outer iterations for ES. It is also observed that the TS algorithm requires more iterations for convergence compared to the ES and MS algorithms, which shows stable performance after around {$10$} iterations. This is expected, as the extra time-domain resource brings in more optimization variables and higher computational complexity, thereby reducing the convergence rate. Nevertheless, all the three algorithms show their convergence within limited number of iterations, which implies the possible application in practice.  

\begin{figure}
	\centering
	\includegraphics[scale=0.6]{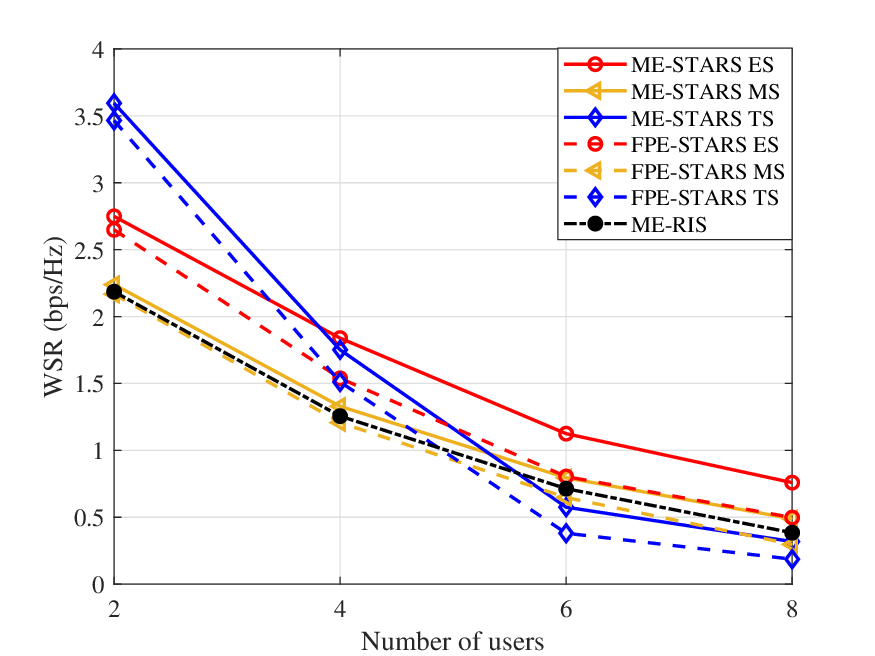}
	\caption{WSR versus number of users.}
	\label{fig:WSR-user-number} 
\end{figure} 
In Fig.~\ref{fig:WSR-user-number}, we compare the achievable WSR of different scehmes versus the number of users $J$. Besides the performance achieved by ME-STARS with three operating protocols, we also provide the results obtained by the benchmark schemes. As can be observed, WSR decreases with the increment of $J$ for all schemes, which is expected as the inter-user interference is increased with larger $J$. Regarding the performance comparison among the three protocols for STARS, it is interesting to find that TS is preferable when there exists only two users (one user in the transmission region and the other one in the reflection region). This is because, TS can realize the interference-free communication for each user, given that only one user is in service for each time instant. However, when $J$ becomes larger than $2$, the multiuser interference is also introduced for TS. Then, ES becomes appealing again since it can make full use of the entire communication resources without the need of time-domain division. ES also shows its superiority compared to MS, which is expected as ES has higher flexibility in beamforming, given that each element can work in transmission and reflection modes simultaneously.

{It is also seen from Fig.~\ref{fig:WSR-user-number} that the gaps between the ME-STARS and the FPE-STARS get enlarged for all protocols when $J$ increases.} The reason is given as follows. 
If $J$ is small, the multiuser interference can be efficiently mitigated by the active and passive beamforming. As $J$ increases to get close to the number of FPAs at the BS or the number of MEs at the STARS, i.e., $M/N$, the interference among users can not be well suppressed due to the highly correlated channels. However, the ME-STARS can reduce the correlations via the MEs position adjustment, thereby improving the WSR performance. Thus, Fig.~\ref{fig:WSR-user-number} verifies that the ME-STARS can facilitate higher performance gain over the FPE-STARS with a large number of users. Regarding the performance comparison between the ME-STARS and the ME-RIS, the ME-STARS shows superior performance for both ES and MS. The reason is that the ME-STARS possesses higher DoFs for enhancing the desired signal power and mitigating inter-user interference for both ES and MS. However, for TS, the ME-STARS shows worse performance than the ME-RIS when $J$ gets large. The reason behind this is that, the full usage of the available communication time of ME-RIS is beneficial for the achievable WSR, especially when the multiuser interference is also non-negligible for TS and the time resource dominates the performance gain.

\begin{figure}
    \centering
        \includegraphics[scale=0.6]{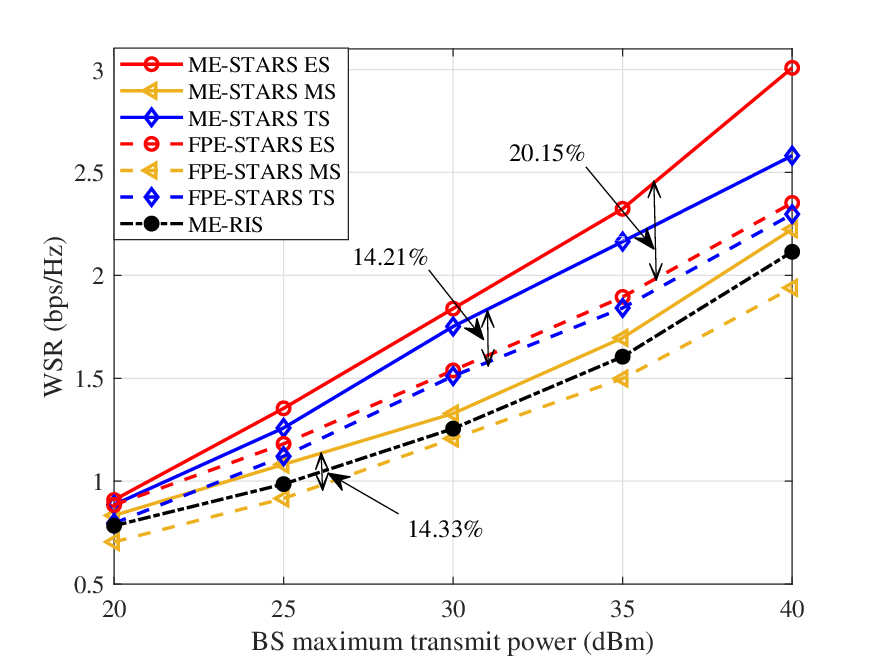}
    \caption{WSR versus BS maximum transmit power.}
    \label{fig:WSR-power}
\end{figure}
In Fig.~\ref{fig:WSR-power}, we investigate the achievable WSR versus the BS maximum transmit power $P^{\text{max}}$. One can observe that, the ME-STARS outperforms the FPE-STARS by around $20.15\%$, $14.33\%$, and $14.21\%$ for ES, MS, and TS, respectively, in average. This observation validates our theoretical analysis that the ME-STARS can further explore the spatial-domain diversity and therefore results in considerable WSR gain. Note that the WSR improvement for the ME-STARS compared to the FPE-STARs is most significant when ES protocol is adopted. This is due to the reason that, ES possesses the highest DoFs for passive beamforming, and thus the MEs positions optimization can lead to greater performance gain. 

\begin{figure}
	\centering
	\includegraphics[scale=0.6]{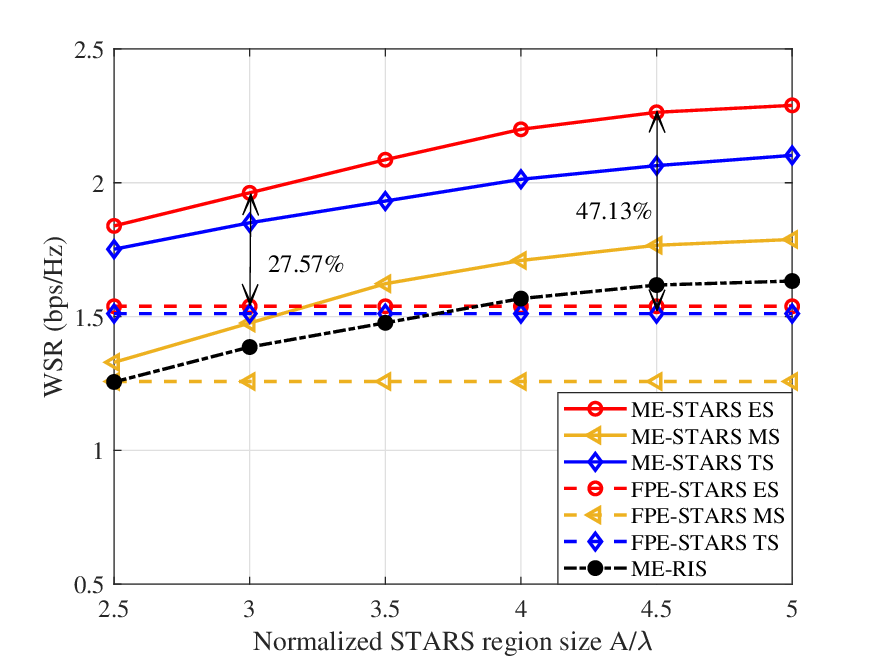}
	\caption{WSR versus normalized ME-STARS region size.}
	\label{fig:WSR-region-size} 
\end{figure} 
    In Fig.~\ref{fig:WSR-region-size}, we demonstrate the achievable WSR versus the ME-STARS normalized movable region size $A/\lambda$. It can be observed that the ME-STARS outperforms the FPE-STARS for all movable region sizes, and the performance gain is more pronounced when the movable region gets larger. {For example, the ME-STARS achieves $27.57\%$ and $47.13\%$ higher WSR compared to the FPE-STARS for ES, when $A/\lambda$ is set as $3$ and $4.5$, respectively.} This is because, higher spatial-domain diversity gain can be explored when the MEs moving region is expanded. It is also shown that the WSR gets saturated with the increment of $A/\lambda$. This is due to the fact that, when there are sufficient optimal positions for the given number of antennas, enlarging the region size cannot bring in any further performance gain. Therefore, we can infer that the maximal WSR can be obtained within a finite ME-STARS movable region size. 

\begin{figure}
	\centering
	\includegraphics[scale=0.6]{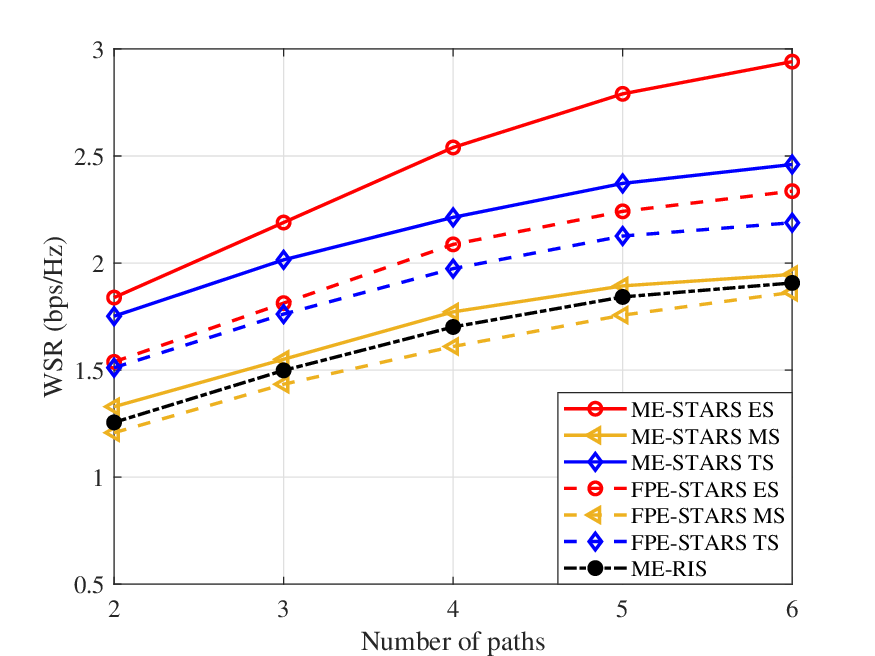}
	\caption{WSR versus number of paths of each channel.}
	\label{fig:WSR-path-number} 
\end{figure} 
In Fig.~\ref{fig:WSR-path-number}, we study the WSR performance versus different numbers of channel paths $L$. It can be observed that the ME-STARS working in the ES protocol achieves the highest WSR across all $L$. Also observe that the achievable WSR for all schemes increases with the increment of $L$. The reason behind this can be explained as follows. On the one hand, the multiuser channel matrix possesses higher rank when $L$ gets larger, which means that the correlations among the users channel vectors are decreased. This is beneficial for mitigating the multiuser interference. On the other hand, since the small-scale fading is pronounced when there exists more multi-path components, the number of optimal MEs positions increases within the given movable region. As such, the MEs schemes can achieve superior performance with more available local maxima.

\begin{figure}
	\centering
	\includegraphics[scale=0.6]{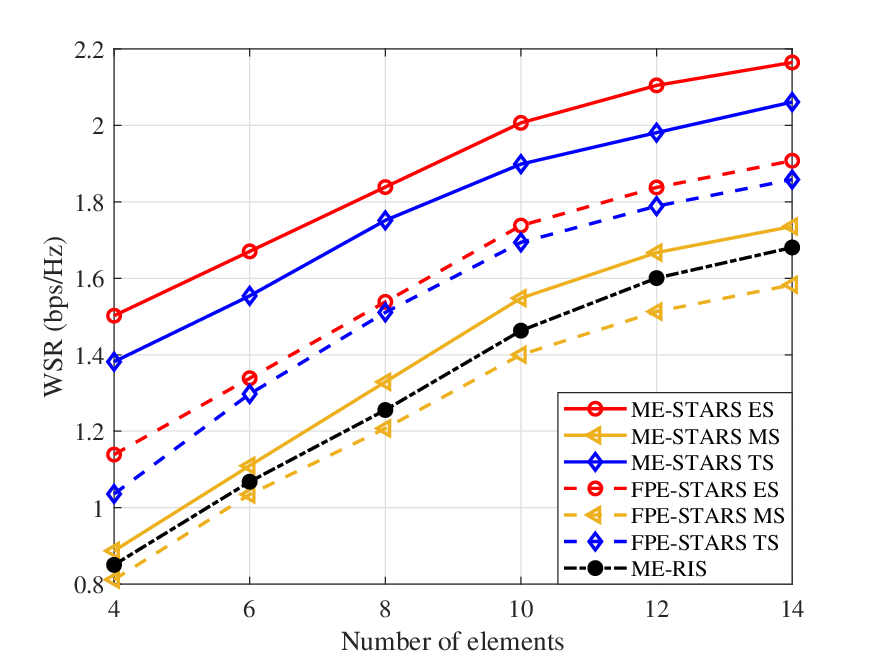}
	\caption{WSR versus number of ME-STARS elements.}
	\label{fig:WSR-element-number} 
\end{figure} 
In Fig.~\ref{fig:WSR-element-number}, we present the impact of the number of elements $N$ on the ME-STARS performance. One can first observe that the WSR obtained by all schemes increases with the increment of $N$, while the ME-STARS achieves the highest WSR for ES. This is expected as larger $N$ enables higher passive beamforming gain, thereby enhancing the overall communication capacity. Moreover, when $N$ increases, the ME-STARS can move more elements to the preferable positions, thus further mitigating the interference and enhancing the desired signals. {It can also be observed that the increasing rate of the MEs schemes gets smaller with the increment of $N$.} This is due to the fact that, since the local optimal positions in the given movable region is limited and the antennas positions are restricted by the minimum distance, not all the elements can be moved to the preferable positions when $N$ becomes larger than the number of local maxima. 

\setcounter{equation}{39} % 当前公式序号变为x，x等于长公式应有的序号减1.
\begin{figure*}[hb]
    \hrulefill
    \begin{align}
    \label{eq:y-r-n}
        &\upsilon_j(\mathbf{u}_n)  \ = \left|\mathbf{a}_{n,j}\mathbf{F}_j(\mathbf{u}_n)\mathbf{b}_j + c_{n,j}\right|^2\nonumber\\
     & \ = \left|\sum_{o=1}^{L_{\text{BS}}}\sum_{p=1}^{L_{\text{S},j}}\left|a_{n,j}^p\right|\left|{b}_j^o\right|e^{j\left[\frac{2\pi}{\lambda}\left(\rho_{\text{S},j}^p(\mathbf{u}_n)-\rho_{\text{S,in}}^o(\mathbf{u}_n)\right)+\angle b_j^o +\angle a_{n,j}^p\right]}+|c_{n,j}|e^{j\angle c_{n,j}}\right|^2\nonumber \\
     & \ = \left(\sum_{o=1}^{L_{\text{BS}}}\sum_{p=1}^{L_{\text{S},j}}\left|a_{n,j}^p\right|\left|{b}_j^o\right|\cos \phi_{j}^{p,o}(\mathbf{u}_n)+\left|c_{n,j}\right|\cos \angle c_{n,j}\right)^2 + \left(\sum_{o=1}^{L_{\text{BS}}}\sum_{p=1}^{L_{\text{S},j}}\left|a_{n,j}^p\right|\left|{b}_j^o\right|\sin \phi_{j}^{p,o}(\mathbf{u}_n)+\left|c_{n,j}\right|\sin \angle c_{n,j}\right)^2.
\end{align}

    \begin{align}
    \label{eq:phi}
    \phi_j^{p,o}(\mathbf{u}_n) = \frac{2\pi}{\lambda}\left[\left(\underbrace{\cos{\theta_{\text{S},j}^p}\sin{\phi_{\text{S},j}^p} - \cos{\theta_{\text{BS,S}}^o}\sin{\phi_{\text{BS,S}}^o}}_{\zeta_j^{p,o}}\right)x_n + \left(\underbrace{\sin{\theta_{\text{S},j}^p}-\sin{\theta_{\text{BS,S}}^o}}_{\mu_j^{p,o}}\right)y_n\right]+\angle b_j^o +\angle a_{n,j}^p.
\end{align}

    \begin{subequations}
    \label{eq:gradient-vector}
        \begin{align}
             \frac{\partial \upsilon_j(\mathbf{u}_n)}{\partial x_n} = \frac{4\pi\left|c_{n,j}\right|}{\lambda}\sum_{o=1}^{L_{\text{BS}}}\sum_{p=1}^{L_{\text{S},j}}\zeta_j^{p,o}|a_{n,j}^p||b_j^o|\sin \left(\angle c_{n,j} - \phi_j^{p,o}(\mathbf{u}_n)\right),
        \end{align}
        \begin{align}
            \frac{\partial \upsilon_j(\mathbf{u}_n)}{\partial y_n} = \frac{4\pi\left|c_{n,j}\right|}{\lambda} \sum_{o=1}^{L_{\text{BS}}}\sum_{p=1}^{L_{\text{S},j}}\mu_j^{p,o}|a_{n,j}^p||b_j^o|\sin \left(\angle c_{n,j} - \phi_j^{p,o}(\mathbf{u}_n)\right).
        \end{align}
    \end{subequations}
\end{figure*}

\section{Conclusion}
In this paper, we proposed the ME-STARS-aided multiuser MISO communication system. Different from the conventional FPE-STARS where the elements positions are fixed, the proposed ME-STARS further exploits the spatial-domain diversity by moving the elements to preferable positions within a confined region. The WSR maximization problems were formulated for three STARS operating protocols, i.e., ES, MS, and TS, where the elements positions and the active/passive beamforming were highly coupled. To solve the resulting non-convex problem for ES, the AO-based algorithm was proposed for decomposing the original problem into three subproblems. Subsequently, the GDA embedded with the penalty method was employed to efficiently search for the local-optimal MEs positions, while the WMMSE and the SCA methods were utilized for solving the active and passive beamforming subproblems, respectively. It was demonstrated that the proposed AO algorithm proposed for ES can be extended to solve the MS and TS problems. Extensive simulation results were provided to verify the superiority of the proposed ME-STARS assisted multiuser communications. It was shown that the ME-STARS can significantly improve the WSR compared to the FPE-STARS for all operating protocols, thanks to the higher spatial-domain diversity. Moreover, the ME-STARS was demonstrated to outperform the ME-RIS for both ES and MS due to the greater flexibility in transmission and reflection beamforming. The ME-STARS also showed its superiority in the scenarios with a large number of users or rich scatterers, for the more pronounced gain in interference mitigation. 

\begin{appendices} 
\section{Derivation of ${\partial R_j^{\text{ES}}}/{\partial \mathbf{u}_n}$}
\label{app:A}
For ease of brevity, define $\upsilon_j(\mathbf{u}_n) = \left|\mathbf{q}_{\kappa}^H\mathbf{V}_{j}\left(\mathbf{u}_n\right)\mathbf{w}_{j}\right|^2$ and $\gamma_j(\mathbf{u}_n) = \sum_{j'\neq j}^J\left|\mathbf{q}_{\kappa}^H\mathbf{V}_{j}\left(\mathbf{u}_n\right)\mathbf{w}_{j'}\right|^2$. Based on the chain rule, we have 
\setcounter{equation}{37}
\begin{equation}
\frac{\partial R_j^{\text{ES}}}{\partial \mathbf{u}_n} = \frac{\frac{\partial \upsilon_j(\mathbf{u}_n)}{\partial \mathbf{u}_n}\left( \gamma_j(\mathbf{u}_n) + \sigma_j^2\right)- \upsilon_j(\mathbf{u}_n) \frac{\partial \gamma_j(\mathbf{u}_n)}{\partial \mathbf{u}_n}}{\ln2 \left( 1 + \upsilon_j(\mathbf{u}_n)/(\gamma_j(\mathbf{u}_n) +\sigma_j^2)\right)\left( \gamma_j(\mathbf{u}_n) +\sigma_j^2\right)^2}.
\label{eq:partial-R-j}
\end{equation}
For the derivation of the gradient vector of $\upsilon_j(\mathbf{u}_n)$ w.r.t. $\mathbf{u}_n$, i.e., $\nabla \upsilon_j(\mathbf{u}_n) = \left[\begin{matrix}
        \frac{\partial \upsilon_j(\mathbf{u}_n)}{\partial x_n} & \frac{\partial \upsilon_j(\mathbf{u}_n)}{\partial y_n}
    \end{matrix}\right]^T$, we first define constant $c_{n,j}$ as
\begin{equation}
    c_{n,j} = \left|c_{n,j}\right|e^{j\angle c_{n,j}} \triangleq \sum_{n'\neq n}^N \mathbf{a}_{n',j}\mathbf{F}_j(\mathbf{u}_{n'})\mathbf{b}_j,
\end{equation}
where $|c_{n,j}|$ and $\angle c_{n,j}$ denote the amplitude and phase of $c_{n,j}$, respectively.
Further denote the $p$-th element of $\mathbf{a}_{n,j}$ and the $o$-th element of $\mathbf{b}_j$ as ${a}_{n,j}^p = \left|a_{n,j}^p\right|e^{j\angle a_{n,j}^p}$ and ${b}_j^o = \left|{b}_j^o\right|e^{j\angle {b}_j^o}$, respectively, with the $a_{n,j}^p$ amplitude $\left|a_{n,j}^p\right|$ and phase $\angle a_{n,j}^p$, as well as the ${b}_j^o$ amplitude $\left|{b}_j^o\right|$ and phase $\angle {b}_j^o$. According to \eqref{eq:element-wise}, $\mathbf\upsilon_j(\mathbf{u}_n)$ can be rewritten as in~\eqref{eq:y-r-n}, where $\phi_j^{p,o}(\mathbf{u}_n)$ is defined as in~\eqref{eq:phi}. Subsequently, $\frac{\partial \upsilon_j(\mathbf{u}_n)}{\partial x_n}$ and $\frac{\partial \upsilon_j(\mathbf{u}_n)}{\partial y_n}$ are given in~\eqref{eq:gradient-vector}.

Similarly, for the derivation of the gradient vector of $\gamma_j(\mathbf{u}_n)$ w.r.t. $\mathbf{u}_n$, i.e., $\nabla \gamma_j(\mathbf{u}_n) = \left[\begin{matrix}
        \frac{\partial \gamma_j(\mathbf{u}_n)}{\partial x_n} & \frac{\partial \gamma_j(\mathbf{u}_n)}{\partial y_n}
    \end{matrix}\right]^T$, we first define constant $i_{n,j}^{j'}$ as
    \setcounter{equation}{42}
\begin{equation}
    i_{n,j}^{j'} = \left|i_{n,j}^{j'}\right|e^{\angle i_{n,j}^{j'}}\triangleq \sum_{n'\neq n}^N\mathbf{a}_{n',j}\mathbf{F}_j(\mathbf{u}_{n'})\mathbf{b}_{j'},
\end{equation}
where $|i_{n,j}^{j'}|$ and $\angle i_{n,j}^{j'}$ denote the amplitude and phase of $i_{n,j}^{j'}$, respectively. Moreover, denote the $o$-th element of $\mathbf{b}_{j'}$ as ${b}_{j'}^o = \left|{b}_{j'}^o\right|e^{j\angle {b}_{j'}^o}$, with the amplitude $\left|{b}_{j'}^o\right|$ and the phase $\angle {b}_{j'}^o$. Then $\gamma_j\left(\mathbf{u}_n\right)$ can be rewritten as in~\eqref{eq:gamma-j-n}, where $\varphi_{j}^{p,o}(\mathbf{u}_n)$ is defined as $\varphi_j^{p,o}(\mathbf{u}_n) = \frac{2\pi}{\lambda}\left({\zeta_j^{p,o}}x_n + {\mu_j^{p,o}}y_n\right)+\angle b_{j'}^o +\angle a_{n,j}^p$.
\begin{figure*}
    \begin{align}
    \label{eq:gamma-j-n}
        & \gamma_j(\mathbf{u}_n) = \sum_{j'\neq j}^J\left|\mathbf{a}_{n,j}\mathbf{F}_j(\mathbf{u}_n)\mathbf{b}_{j'} + i_{n,j}^{j'}\right|^2\nonumber\\
     & = \sum_{j'\neq j}^J\left|\sum_{o=1}^{L_{\text{BS}}}\sum_{p=1}^{L_{\text{S},j}}\left|a_{n,j}^p\right|\left|{b}_{j'}^o\right|e^{j\left[\frac{2\pi}{\lambda}\left(\rho_{\text{S},j}^p(\mathbf{u}_n)-\rho_{\text{S,in}}^o(\mathbf{u}_n)\right)+\angle b_{j'}^o +\angle a_{n,j}^p\right]}+|i_{n,j}^{j'}|e^{j\angle i_{n,j}^{j'}}\right|^2\nonumber \\
     & = \sum_{j'\neq j}^J\left[\left(\sum_{o=1}^{L_{\text{BS}}}\sum_{p=1}^{L_{\text{S},j}}\left|a_{n,j}^p\right|\left|{b}_{j'}^o\right|\cos \varphi_{j}^{p,o}(\mathbf{u}_n)+\left|i_{n,j}^{j'}\right|\cos \angle i_{n,j}^{j'}\right)^2 + \left(\sum_{o=1}^{L_{\text{BS}}}\sum_{p=1}^{L_{\text{S},j}}\left|a_{n,j}^p\right|\left|{b}_{j'}^o\right|\sin \varphi_{j}^{p,o}(\mathbf{u}_n)+\left|i_{n,j}^{j'}\right|\sin \angle i_{n,j}^{j'}\right)^2\right].
\end{align}
\end{figure*}
Subsequently, $\frac{\partial \gamma_j(\mathbf{u}_n)}{\partial x_n}$ and $\frac{\partial \gamma_j(\mathbf{u}_n)}{\partial y_n}$ are given in~\eqref{eq:gamma-gradient-vector}.
\begin{figure*}
    \begin{subequations}
    \label{eq:gamma-gradient-vector}
        \begin{align}
             \frac{\partial \gamma_j(\mathbf{u}_n)}{\partial x_n} = \frac{4\pi}{\lambda} \sum_{j'\neq j}^J\sum_{o=1}^{L_{\text{BS}}}\sum_{p=1}^{L_{\text{S},j}}\zeta_j^{p,o}\left|i_{n,j}^{j'}\right||a_{n,j}^p||b_{j'}^o|\sin\left({\angle i_{n,j}^{j'} - \varphi_j^{p,o}(\mathbf{u}_n)}\right),
        \end{align}
        \begin{align}
            \frac{\partial \gamma_j(\mathbf{u}_n)}{\partial y_n} = \frac{4\pi}{\lambda} \sum_{j'\neq j}^J\sum_{o=1}^{L_{\text{BS}}}\sum_{p=1}^{L_{\text{S},j}}\mu_j^{p,o}\left|i_{n,j}^{j'}\right||a_{n,j}^p||b_{j'}^o|\sin\left({\angle i_{n,j}^{j'} - \varphi_j^{p,o}(\mathbf{u}_n)}\right).
        \end{align}
    \end{subequations}
    \hrulefill
\end{figure*}

\end{appendices}

\bibliographystyle{IEEEtran}
\bibliography{mybib}
\end{document}